\documentclass[11pt,a4paper]{article} 
\usepackage{jheppub}

\usepackage{graphicx}
\usepackage{amsmath}
\usepackage{amsfonts}
\newcommand{\beq}{\begin{equation}}
\newcommand{\eeq}{\end{equation}}
\newcommand{\beqa}{\begin{eqnarray}}
\newcommand{\eeqa}{\end{eqnarray}}
\def\lsim{\raise0.3ex\hbox{$<$\kern-0.75em\raise-1.1ex\hbox{$\sim$}}}
\def\gsim{\raise0.3ex\hbox{$>$\kern-0.75em\raise-1.1ex\hbox{$\sim$}}}

\def\0{{\bf 0}}

\def\kk{{\kappa}}
\def\pp{{\hat{p}}}

\title{Development of heavy-flavour flow-harmonics in high-energy nuclear collisions}

\author[a]{Andrea Beraudo}
\author[a]{Arturo De Pace}
\author[a]{Marco Monteno}
\author[a]{Marzia Nardi}
\author[a]{Francesco Prino}

\affiliation[a]{INFN, Sezione di Torino, via Pietro Giuria 1, I-10125 Torino}

\emailAdd{beraudo@to.infn.it}
\emailAdd{depace@to.infn.it}
\emailAdd{monteno@to.infn.it}
\emailAdd{nardi@to.infn.it}
\emailAdd{prino@to.infn.it}

\abstract{We employ the POWLANG transport setup, developed over the last few years, to provide new predictions for several heavy-flavour observables in relativistic heavy-ion collisions from RHIC to LHC center-of-mass energies. In particular, we focus on the development of the flow-harmonics $v_2$ and $v_3$ arising from the initial geometric asymmetry in the initial conditions and its associated event-by-event fluctuations. Within the same transport framework, for the sake of consistency, we also compare the nuclear modification factor of the $p_T$ spectra of charm and beauty quarks, heavy hadrons and their decay electrons. We compare our findings to the most recent data from the experimental collaborations. We also study in detail the contribution to the flow harmonics from the quarks decoupling from the fireball during the various stages of its evolution: although not directly accessible to the experiments, this information can shed light on the major sources of the final measured effect.}

\begin{document}

\maketitle

\section{Introduction}
Experimental measurements of heavy-flavour production in relativistic heavy-ion collisions are a major tool to get information on the properties of the (deconfined, if a sufficiently high energy-density is achieved) medium formed in these events, in particular on its transport coefficients~\cite{Adare:2006nq,ALICE:2012ab,Abelev:2013lca,Chatrchyan:2012np,Adamczyk:2014uip,Adamczyk:2017xur,Sirunyan:2017plt}. At high momentum the major effect of the interaction with the medium is a quenching of the heavy-quark momentum spectra due to parton energy-loss: this provides information on the medium opacity~\cite{ALICE:2012ab,Chatrchyan:2012np}.  
At low/intermediate momenta, on the other hand, if the transport coefficients were large enough, heavy quarks would even approach local thermal equilibrium with the rest of the medium, taking part in its collective expansion~\cite{Moore:2004tg}. This would lead to clear signatures in the final observables: the radial and elliptic flow of the fireball arising from the heavy-ion collision would leave their fingerprints also in the heavy-flavour sector, boosting the heavy quarks from low to moderate momenta and giving rise to azimuthal anisotropies in their angular distributions~\cite{Abelev:2013lca,Adamczyk:2014uip}. Furthermore, since nowadays higher flow-harmonics ($v_3,v_4,v_5...$) of soft-hadron azimuthal distributions are measured (providing information on event-by-event fluctuations and granularity of the initial conditions), one would like to address this issue also in the heavy-flavour sector (for first experimental results, see Ref.~\cite{Sirunyan:2017plt}): this will be one of the major topics dealt with in this paper, based on the POWLANG transport setup developed by the authors over the last years~\cite{Alberico:2011zy,Alberico:2013bza,Beraudo:2014boa,Beraudo:2015wsd}. A similar theoretical study was performed in~\cite{Nahrgang:2014vza} and, accounting only for the path-length dependence of parton energy-loss, in~\cite{Prado:2016vgz}. The long term goal would be to perform this kind of study on an event-by-event basis selecting, within the same centrality class, collisions characterized by different initial eccentricities or comparing events from different centrality classes but having a comparable initial eccentricity: we believe that this has the potential to further constrain the coupling of the heavy quarks with the background medium. However, the starting point is to check to be able to reproduce the trend of the experimental data in the case of the event-averaged results, which is the subject of the present paper.  

Of course, non-trivial features in the heavy-flavour hadronic distributions experimentally measured can not be directly ascribed to the parent heavy quarks. As suggested by several studies, in the presence of a deconfined medium, rather then fragmenting like in the vacuum, heavy quarks may hadronize by recombining with the light thermal partons nearby to give rise to open charm/beauty hadrons. Belonging to the non-perturbative realm of QCD, there is no solid first-principle theory to describe hadronization, neither in the vacuum nor in the medium. The latter is modeled in several different ways in the literature -- via coalescence~\cite{Greco:2003vf,Gossiaux:2009mk,Nahrgang:2016lst}, formation of color-singlet clusters/strings~\cite{Beraudo:2014boa} or of resonances~\cite{Ravagli:2007xx,He:2011qa,He:2014cla} -- but the qualitative effect is always the same: the light thermal quark involved in the recombination process is part of a fluid cell sharing a common collective velocity and this provides an additional contribution to the (radial, elliptic and also triangular, as will be shown in the paper) flow of the final heavy-flavour hadron. Clearly, recombination with light partons from the medium, besides the kinematic distributions, can also affect the heavy-flavour hadrochemistry in nucleus-nucleus collisions, changing the relative yields of the various species with respect to the proton-proton case. This was modeled for instance in~\cite{He:2014cla} and first experimental results are getting available~\cite{Adam:2015jda,ALICE-PUBLIC-2017-003,Zhou:2017ikn}, however we will not touch such an issue.

%As already said, there is evidence that an important fraction of the radial, elliptic and also triangular flow of heavy-flavour particles arises from hadronization, through the recombination with light partons from the surrounding deconfined medium.
Concerning the flow acquired by charm and beauty quarks during the partonic phase it is of interest to disentangle the various sources of possible azimuthal anisotropies, in order to better understand how much heavy quarks really approach thermal equilibrium, tending to flow with the fireball, and how much of the final signal instead is simply due to trivial geometric effects. We will address this issue by studying the temporal development of the elliptic and triangular flow, disentangling the contribution to the final $v_2$ and $v_3$ from the heavy quarks decoupling at different times. A somehow similar analysis, referring to the bulk soft-particle production, was performed in~\cite{He:2015hfa} where the elliptic and triangular flow were studied within a transport model as a function of the number of collisions suffered by the partons; the authors found that the anisotropic escape probability of the partons, trivially arising from the initial geometry, provides a major contribution to the final signal, challenging the usual hydrodynamic interpretation of the data based on the formation of a strongly-interacting medium.
Actually, our analysis deals only with the propagation of heavy quarks and is based on a more macroscopic approach, since the background medium is given a coarse-grained hydrodynamic description and the propagation of the heavy quarks throughout the fireball is not modeled through the individual collisions with the other partons, but just in terms -- according to the Langevin equation -- of an average squared-momentum exchange per unit time. Our findings will be presented and discussed in a devoted section; here we only anticipate that the final result come from a non-trivial interplay of contributions from the heavy quarks decoupling during all the stages of the fireball evolution.

Our paper is organized as follows. In Sec.~\ref{sec:transport} we present a detailed description of the transport equations implemented in the POWLANG setup. In Sec.~\ref{sec:medium} we describe how we model the initialization and the evolution of the background medium, in particular in the case of fluctuating initial conditions giving rise to a triangular flow. In Sec.~\ref{sec:flow-development} we study the temporal development of the heavy-quark $v_2$ and $v_3$. In Sec.~\ref{sec:results} POWLANG results for various heavy-flavour observables in nucleus-nucleus collisions are compared to recent experimental results obtained at RHIC and at the LHC. Finally, in Sec.~\ref{sec:discussion} we summarize the main conclusions of our paper and outline possible future developments of our studies.
  
%Since~\cite{Beraudo:2014boa,Rapp:2008qc,He:2015hfa,Heiselberg:1998es,Kolb:2000fha,Nahrgang:2014vza,Adamczyk:2017xur,CMS-PAS-HIN-16-007}
\section{The transport setup}\label{sec:transport}
Different approaches are adopted in the literature to model the heavy-flavour transport throughout the plasma of light quarks and gluons expected to be produced in heavy-ion collisions.
The POWLANG setup is based on the relativistic Langevin equation; for the latter different implementations can be found and this can be sometimes a source of confusion. Hence, here we briefly summarize the essential points of our transport scheme.

The starting point of any transport calculation is the Boltzmann equation for the evolution of the heavy-quark phase-space distribution 
\beq
\frac{d}{dt}f_Q(t,\vec x,\vec p)=C[f_Q]\quad{\rm with}\quad C[f_Q]=\int\!\! d\vec q\,[{w(\vec p+\vec q,\vec q)f_Q(t,\vec x,\vec p+\vec q)}-{w(\vec p,\vec q)f_Q(t,\vec x,\vec p)}],\label{eq:Boltzmann}
\eeq
where the collision integral $C[f_Q]$ is expressed in terms of the $\vec p\to\vec p-\vec q$ transition rate $w(\vec p,\vec q)$. The direct solution of the Boltzmann integro-differential equation is numerically demanding (for a detailed description of the approach see for instance~\cite{Xu:2004mz}); however, as long as $q\ll p$ (in a relativistic gauge plasma $q$ is typically of order $gT$, $g$ being the QCD coupling and $T$ the temperature), one can expand the collision integrand in powers of the momentum exchange. Truncating the expansion to second order corresponds to the Fokker-Planck (FP) approximation, which, for a homogeneous and isotropic system, gives
\beq
\frac{\partial}{\partial t}f_Q(t,\vec p)=\frac{\partial}{\partial p^i}\left\{{A^i(\vec p)}f_Q(t,\vec p)+\frac{\partial}{\partial p^j}[{B^{ij}(\vec p)}f_Q(t,\vec p)]\right\},
\eeq
where
{\setlength\arraycolsep{1pt}
\beqa
{A^i(\vec p)}=\int d\vec q\, q^iw(\vec p,\vec q)\;\;&\longrightarrow&\;\;{A^i(\vec p)={A(p)}\,p^i}\nonumber\\
{B^{ij}(\vec p)}=\frac{1}{2}\int d\vec q\, q^iq^jw(\vec p,\vec q)\;\;&\longrightarrow&\;\;{B^{ij}(\vec p)=(\delta^{ij}-\hat{p}^i\hat{p}^j){B_0(p)}+\hat{p}^i\hat{p}^j{B_1(p)}}.\nonumber
\eeqa}
The study of the heavy-quark propagation in the medium is then reduced to the evaluation of three transport coefficients expressing the friction -- $A(p)$ -- and the momentum-broadening along the transverse and longitudinal directions -- $B_{0/1}(p)$ -- suffered in the plasma. Actually, since one must enforce the asymptotic approach to thermal equilibrium with the medium, the above coefficients (in principle all derived from the scattering matrix) cannot be taken as independent, but are related by the Einstein fluctuation-dissipation relation
\beq
{A(p)=\frac{B_1(p)}{TE_p}-\left[\frac{1}{p}\frac{\partial B_1(p)}{\partial p}+\frac{d-1}{p^2}(B_1(p)-B_0(p))\right]},\label{eq:Einstein}
\eeq
which establishes a link between the momentum broadening and the friction force felt by the heavy quark ($d$ being the number of spatial dimensions). Our choice, in the POWLANG setup, is to evaluate $B_0(p)$ and $B_1(p)$ from first principles and to get $A(p)$ from Eq.~(\ref{eq:Einstein}). 

In order to embed the study of the heavy-quark transport into a setup including the simulation of the initial $Q\overline Q$ production through a pQCD event-generator and the modeling of the evolution of the background medium through a hydrodynamic calculation, it is more convenient to rephrase the FP equation in the form of a discretized Langevin equation:
\beq
{\Delta \vec{p}}/{\Delta t}=-{\eta_D(p)\vec{p}}+{\vec\xi(t)}.\label{eq:Langevin}
\eeq
One no longer deals with the time evolution of a phase-space distribution but rather with the one of a (large) sample of relativistic particles. Eq.~(\ref{eq:Langevin}) provides a recipe to update the heavy quark momentum in the time-step $\Delta t$ through the sum of a deterministic friction force and a random noise term specified by its temporal correlator
\beq
\langle\xi^i(\vec p_t)\xi^j(\vec p_{t'})\rangle\!=\!{b^{ij}(\vec p_t)}{\delta_{tt'}}/{\Delta t}\qquad{b^{ij}(\vec p)}\!\equiv\!{\kk_\|(p)}\pp^i\pp^j+{\kk_\perp(p)}(\delta^{ij}\!-\!\pp^i\pp^j).
\eeq
It can be shown that there is a one-to-one correspondence between the transport coefficients entering into the Langevin equation and the FP ones: $\kappa_\perp(p)\!=\!2B_0(p)$ and $\kk_\|(p)\!=\!2B_1(p)$. Concerning the friction term, the momentum-dependence of the noise-noise correlator (multiplicative noise) requires to consider carefully the discretization of the equation. In the pre-point Ito scheme, in updating the heavy-quark position and momentum during the time-step $t\!\to\! t+\Delta t$, the transport coefficients are evaluated at time $t$ and one can show that, in this case, the friction term coincides with the FP one, $\eta_D(p)\!=\!A(p)$, given in Eq.~(\ref{eq:Einstein}). Other schemes are sometimes employed in the literature: for an overview we refer the reader to Ref.~\cite{He:2013zua}. Since the heavy quarks propagate throughout an expanding fireball, the evaluation of the transport coefficients and the update of their momentum has to be performed at each time-step in the local rest-frame of the fluid, eventually boosting back the result to the laboratory frame~\cite{Alberico:2011zy}, in which the medium flows with four-velocity $u^\mu$.

If the coupling with the background medium were sufficiently strong, the heavy quarks would tend to approach kinetic equilibrium with the plasma in its local rest frame and hence, after boosting to the laboratory frame, to share its collective hydrodynamic flow. 
Within the Langevin setup the interaction between the heavy quark and the medium is summarized (thanks to the Einstein relation) in just two transport coefficients, $\kappa_\perp$ and $\kappa_\|$, which reduces to a single one, $\kappa$, in the non relativistic limit. Theoretical calculations for $\kappa$ in hot-QCD exist in the $M\to\infty$ static-quark limit. Lattice-QCD calculations for the case of a gluon plasma were performed, for various temperatures, in~\cite{Banerjee:2011ra} and recently first continuum-extrapolated results have become available~\cite{Francis:2015daa}, although extracting real-time information from simulations in a Euclidean spacetime introduces large systematic uncertainties. Furthermore, NLO analytic weak-coupling calculations for $\kappa$ were performed in~\cite{CaronHuot:2008uh}, introducing large positive corrections with respect to the tree-level result. Unfortunately the kinematic range in which the most solid theoretical results for $\kappa$ are (so far) available is not the one of relevance for describing (or extracting information from) the experimental data, referring mainly to heavy-flavour particles in a relativistic regime. Hence, in our simulations with weak-coupling transport coefficients, we have to account for their full momentum dependence, evaluating $\kappa_\perp(p)$ and $\kappa_\|(p)$ within a tree-level calculation with Hard Thermal Loop (HTL) resummation of medium effects in the case of interactions mediated by the exchange of soft gluons. In the case of lattice-QCD transport coefficients, on the other hand, no information on the momentum dependence is available and we simply take $\kappa$ from the static calculation of Ref.~\cite{Banerjee:2011ra}, which covers the largest range of temperatures. Actually, an alternative strategy could consist in exploiting the experimental data on heavy-flavour observables (e.g. the nuclear modification factor and the elliptic flow) to estimate \emph{a posteriori} the most probable value of the heavy quark transport coefficients. This was done for charm through a Bayesian analysis in~\cite{Xu:2017hgt}, obtaining results compatible with lattice-QCD calculations.

%%%%%%%%%%%%%%%%%%%%%%%%%%%%%%%%%%%%%%%%%%%%%%  
\section{Modeling of the background medium}\label{sec:medium}
In order to simulate the heavy quark transport in the fireball produced in heavy-ion collisions one needs to model the initial conditions and the subsequent hydrodynamic expansion of the background medium. The initial state is simply taken from the Glauber model, either in its optical or Monte Carlo implementation. As in our past publications~\cite{Alberico:2011zy,Alberico:2013bza,Beraudo:2014boa,Beraudo:2015wsd}, the system is initialized via the entropy-density at the longitudinal proper-time $\tau_0$ ranging, depending on the center-of-mass energy of the collision, from $\tau_0\!=\!1$ fm/c at $\sqrt{s_{\rm NN}}\!=\!200$ GeV to $\tau_0\!=\!0.5$ fm/c at $\sqrt{s_{\rm NN}}\!=\!5.02$ TeV. The hydrodynamic equations describing its evolution are solved through the ECHO-QGP code~\cite{DelZanna:2013eua} in 2+1 dimensions, assuming longitudinal boost-invariance, which is a reasonable approximation to describe observables around mid-rapidity\footnote{We have verified in a few cases that full (3+1)D simulations, numerically more expensive, display negligible differences around mid-rapidity. They will be the subject of a forthcoming publication.}.    

\begin{figure}[!ht]
\begin{center}
\includegraphics[clip,width=0.32\textwidth]{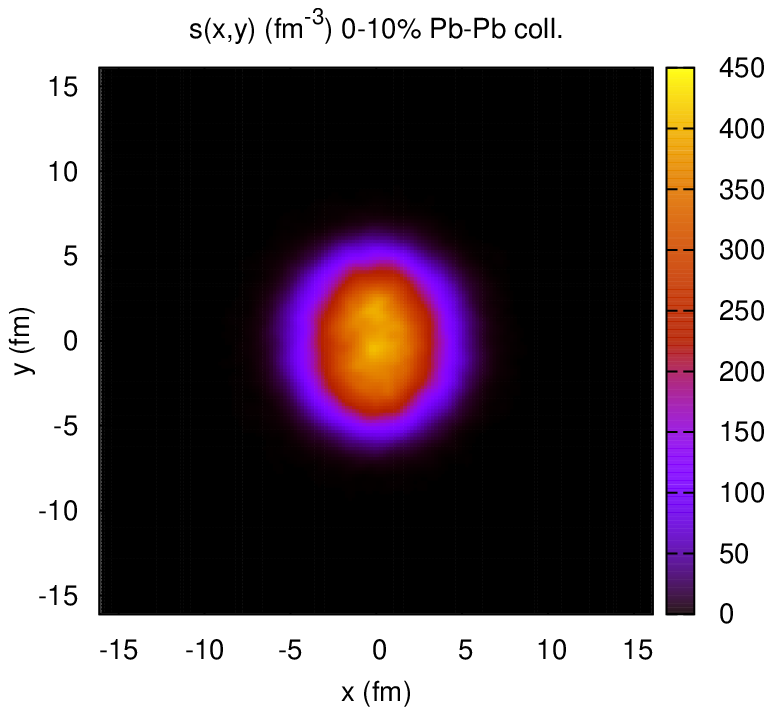}
\includegraphics[clip,width=0.32\textwidth]{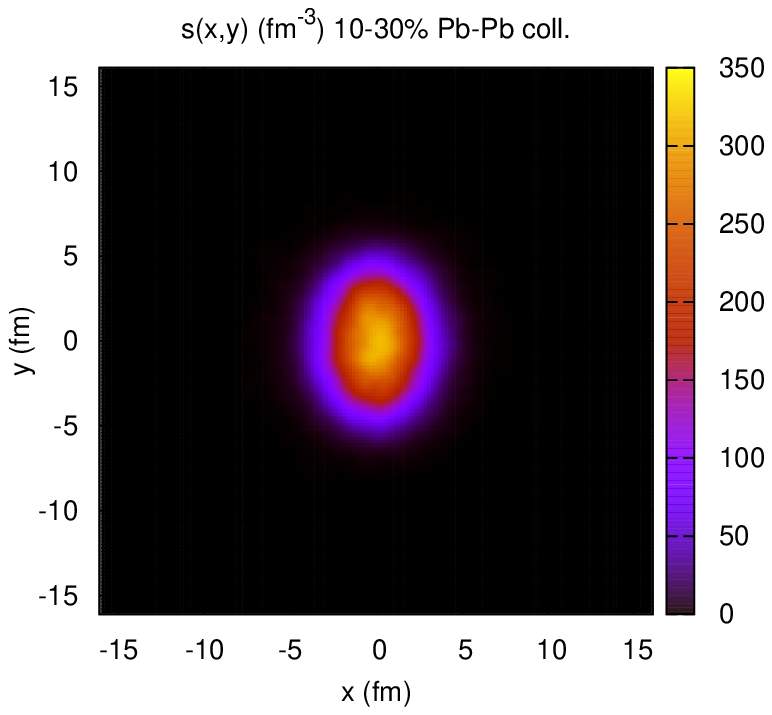}
\includegraphics[clip,width=0.32\textwidth]{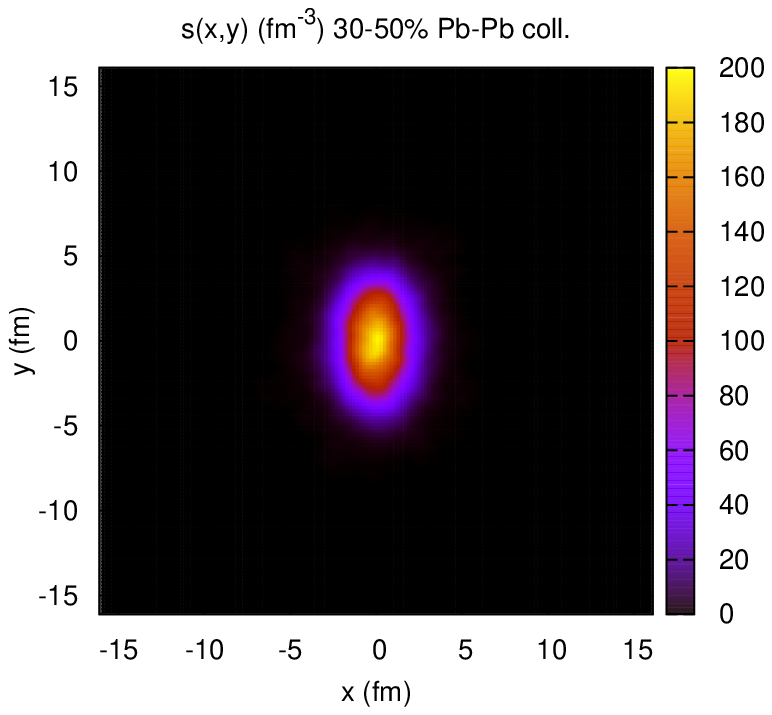}
\caption{Glauber-MC initial conditions for the entropy density at $\tau_0\!=\!0.5$ fm/c for the study of the elliptic flow in Pb-Pb collisions at $\sqrt{s_{\rm NN}}\!=\!5.02$ TeV in different centrality classes. All the events have been rotated to have the event-plane angle $\psi_2$ aligned along the $x$-axis.}\label{fig:eps2MC} 
\end{center}
\end{figure}
\begin{figure}[!ht]
\begin{center}
\includegraphics[clip,width=0.32\textwidth]{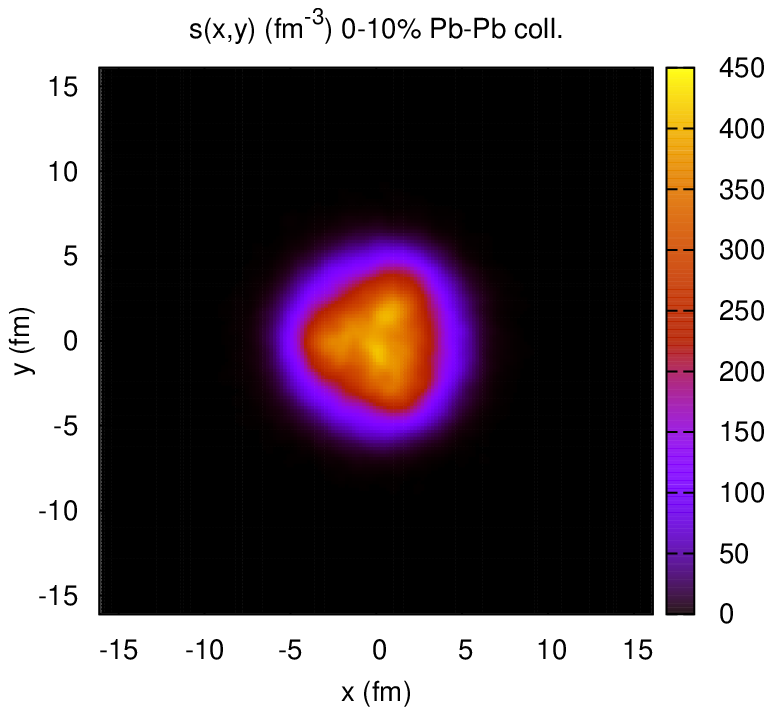}
\includegraphics[clip,width=0.32\textwidth]{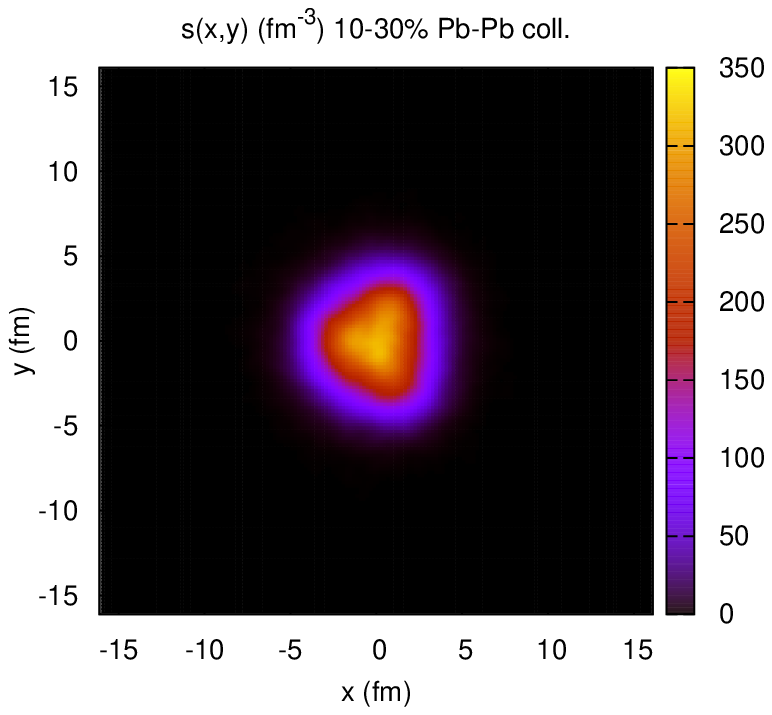}
\includegraphics[clip,width=0.32\textwidth]{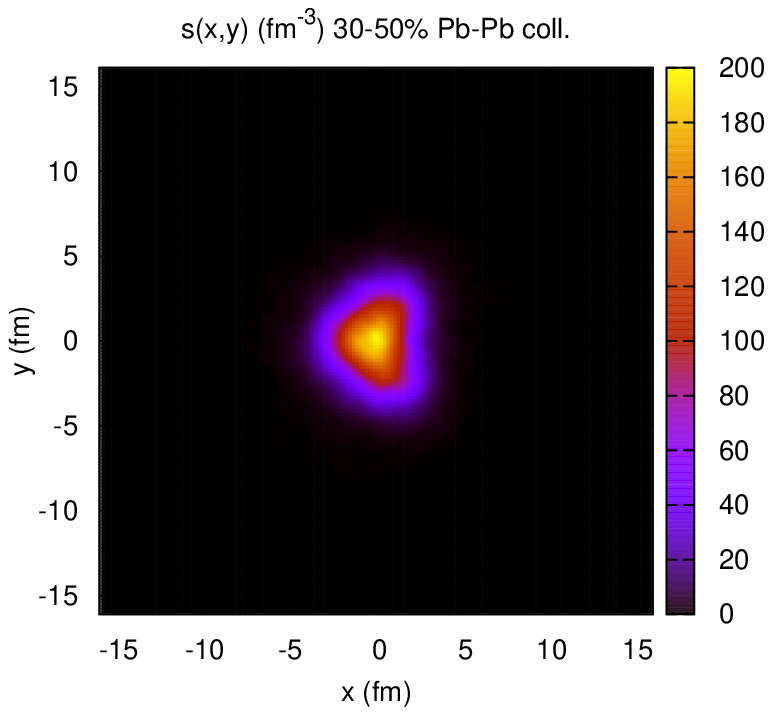}
\caption{Glauber-MC initial conditions for the entropy density at $\tau_0\!=\!0.5$ fm/c for the study of the triangular flow in Pb-Pb collisions at $\sqrt{s_{\rm NN}}\!=\!5.02$ TeV in different centrality classes. All the events have been rotated to have the event-plane angle $\psi_3$ aligned along the $x$-axis.}\label{fig:eps3MC} 
\end{center}
\end{figure}
In the case of a smooth optical-Glauber initialization, as described in~\cite{Alberico:2011zy,Alberico:2013bza,Beraudo:2014boa}, the entropy density at $\tau_0$ is taken as proportional to the local density of binary nucleon-nucleon collisions
\beq
s(x,y|b)=s_0\frac{n_{\rm coll}(x,y|b)}{n_{\rm coll}(0,0|0)},
\eeq
with the parameter $s_0$ fixed by the final rapidity density of charged hadrons at the various center-of-mass energies, from $s_0\!=\!84$ fm$^{-3}$ at $\sqrt{s_{\rm NN}}\!=\!200$ GeV to $s_0\!=\!400$ fm$^{-3}$ at $\sqrt{s_{\rm NN}}\!=\!5.02$ TeV.

For observables like the nuclear modification factor and the elliptic flow  in non-central nucleus-nucleus collisions the optical-Glauber model is sufficient to capture the relevant features of the initial conditions driving the medium evolution. On the other hand, for the study of observables arising from event-by-event fluctuations of the initial geometry like the triangular flow, this is not enough: smooth initial conditions would lead to $v_3\!=\!0$ for any impact parameter of the colliding nuclei and only the granularity of the initial condition can give rise to a non-vanishing triangular flow. Here, as done in~\cite{Beraudo:2015wsd}, we assume that the above lumpiness arises mainly from event-by-event fluctuations in the positions of the nucleons inside the colliding nuclei. We proceed as follows, generalizing to the nucleus-nucleus case the Monte Carlo approach adopted in~\cite{Beraudo:2015wsd} for proton(deuteron)-nucleus collisions. We generate several thousands ($\sim 6000$) of Pb-Pb collisions at random impact parameter and we organize them in centrality classes according to the number of binary nucleon-nucleon collisions. For a given event each nucleon-nucleon collision is taken as a source of entropy production, so that, employing a Gaussian smearing (with $\sigma\!=\!0.2$ fm), we have for the initial entropy density in the transverse plane
\beq
s(x,y)=\frac{K}{2\pi\sigma^2}\sum_{i=1}^{N_{\rm coll}}\exp\left[-\frac{(x-x_i)^2+(y-y_i)^2}{2\sigma^2}\right].
\eeq
For each event the above entropy density can be used as a weight to define complex eccentricities, which characterize the initial state (i.e. both the amount of anisotropy and its orientation in the transverse plane) and are mapped into the final hadron distributions by the hydrodynamic evolution~\cite{Qiu:2011iv}:
\beq
{\epsilon_m}e^{im{\Psi_m}}\equiv-\frac{\left\{r_\perp^2e^{im\phi}\right\}}{\{r_\perp^2\}},\quad{\rm with}\quad\{...\}\equiv\int \!d^2r_\perp\, s(\vec r_\perp)(...).
\eeq
Modulus and orientation of the various azimuthal harmonics are given by:
\beqa
\epsilon_{m}&=&\frac{\sqrt{\{r_\perp^2\cos(m\phi)\}^2+\{r_\perp^2\sin(m\phi)\}^2}}{\{r_\perp^2\}}\\%=-\frac{\{r_\perp^2\cos[m(\phi-\Psi_{m})]\}}{\{r_\perp^2\}}\\
\Psi_{m}&=&\frac{1}{m}\,{\rm atan2}\left(-\{r_\perp^2\sin(m\phi)\},-\{r_\perp^2\cos(m\phi)\}\right)
\eeqa
Exploiting the fact that on an event-by-event basis one has {$v_m\sim\epsilon_m$} for the lowest-order harmonics $m=2,3$, one can consider an {\emph{average background}} obtained through an average of all the events of a given centrality class, each one properly rotated to have the reference angle $\psi_m$  (with $m$ depending on the harmonic being considered) aligned along the $x$-axis and weighted by the number of binary nucleon-nucleon collisions ($Q\overline Q$ production scales according to $N_{\rm coll}$). We applied the above procedure to model the initial conditions of Pb-Pb collisions at $\sqrt{s_{\rm NN}}\!=\!5.02$ TeV, with the purpose of studying within a consistent setup both the elliptic and the triangular flow of heavy-flavour particles after their propagation throughout the medium. As in Ref.~\cite{Beraudo:2015wsd} the contribution to entropy production by each nucleon-nucleon collision was fixed via a matching to an optical-Glauber calculation at the same center-of-mass energy, obtaining $K\tau_0\!=\!6.37$ with an initialization time $\tau_0\!=\!0.5$ fm/c. The resulting initial entropy-density profiles in the transverse plane are displayed, for different centrality classes, in Figs.~\ref{fig:eps2MC} and~\ref{fig:eps3MC}. Notice that, being the angles $\psi_2$ and $\psi_3$ essentially uncorrelated, one gets average initial conditions displaying an almost perfect elliptic/triangular eccentricity.  
%%%%%%%%%%%%%%%%%%%%%%%%%%%%%%%%%%%%%%%%%%%%%%
\section{Development of the heavy-quark flow}\label{sec:flow-development}
The mass-dependent flattening of the hadron $p_T$-spectra observed in relativistic heavy-ion collisions as well as the azimuthal anisotropy of their angular distributions, parametrized in terms of various harmonic coefficients ($v_2$, $v_3$, $v_4$...), have been interpreted for long as signatures of the formation of a strongly interacting medium undergoing a hydrodynamic expansion which, via pressure gradients, translates the initial spatial anisotropy of the system into the final momentum distribution of the particles decoupling from the fireball (for a recent review, see e.g.~\cite{Heinz:2013th}). More and more observables have been analyzed which can be accommodated within a hydrodynamic description like higher flow-harmonics~\cite{ALICE:2011ab}, event-by-event flow fluctuations~\cite{Aad:2013xma} and non-linear effects like interference between different flow-harmonics~\cite{Aad:2015lwa,ALICE:2016kpq}. Notice that, within a kinetic description, in order for a system to behave as a fluid the mean-free-path of its constituents has to be much smaller than the system size, $\lambda_{\rm mfp}\ll L$. The above condition is only marginally satisfied with perturbative partonic cross-sections and hence the idea of the formation of a strongly-interacting QGP was proposed. A further surprise came in the last few years from the observation of analogous effects (mass-dependent radial, elliptic and triangular flow) also in small systems, like the ones produced in high-multiplicity deuteron-nucleus, proton-nucleus and even proton-proton collisions~\cite{Adare:2013piz,Khachatryan:2015waa,Aaboud:2017acw,Khachatryan:2016txc}:  in light of the small size of the medium this makes the hydrodynamic interpretation of the experimental measurements in these events quite challenging and alternative explanations have been proposed (see e.g.~\cite{Blok:2017pui,Dusling:2017aot,Dusling:2017dqg}).

Recently some authors proposed a different paradigm to interpret the above experimental observations. Employing a transport setup with relatively mild partonic cross section of a few mb, they identified the major source of elliptic and triangular flow in their model in the anisotropic escape probability of the partons which decouple from the medium with no or very few interactions, getting a non-vanishing $v_2$ even in the case of small medium opacity~\cite{He:2015hfa}. Similar analytic estimates based on kinetic theory were performed in the past in order to explain the elliptic flow in peripheral nucleus-nucleus collisions in which one expects to produce a less dense medium~\cite{Heiselberg:1998es,Kolb:2000fha}. In~\cite{He:2015hfa}, however, the authors aim at delivering a much stronger message, suggesting that the above mechanism can account for most of the observed effect and questioning the picture of the formation of a strongly-interacting medium, with a collective flow arising from multiple collisions.
     
\begin{figure}[!ht]
\begin{center}
\includegraphics[clip,height=5.8cm]{NQ-vs-FO_comp_10-30}
\includegraphics[clip,height=5.8cm]{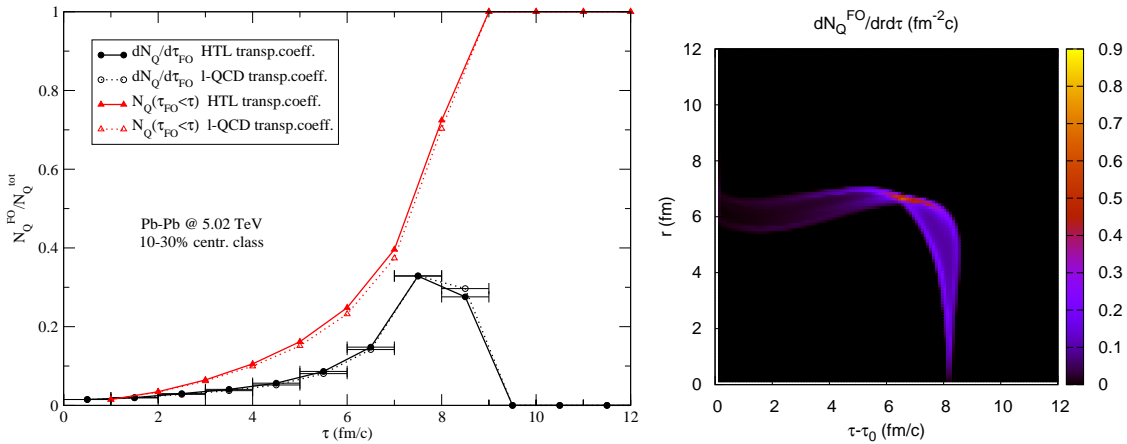}
\caption{Left panel: the decoupling-time distribution of charm quarks escaping from the fireball produced in semi-central Pb-Pb collisions at $\sqrt{s_{\rm NN}}\!=\!5.02$ TeV for different choices of transport coefficients. Right panel: the $\tau-r$ correlation of charm quarks decoupling from the fireball for the same centrality class and HTL transport coefficients.}\label{fig:decoupling} 
\end{center}
\end{figure}
%%%%%%%%%%%%%%%%%%%%%%%%%%
Although the above considerations mainly refer to the bulk particle production, dominated by soft, light hadrons, it is of interest to perform a similar analysis with our heavy-flavour transport model, studying the decoupling of the charm quarks from the fireball (schematically assumed to occur at a temperature $T_{\rm dec}\!=\!155$ MeV) during the various stages of its evolution and how they separately contribute to the anisotropies (elliptic and triangular) of their final (time-integrated) angular distribution.
For an independent and somehow similar analysis, focused mostly on the different time-development of the heavy-flavour $R_{\rm AA}$ and $v_2$, see Ref.~\cite{Rapp:2008qc}.
At variance with a kinetic calculation based on the Boltzmann equation, in which it is possible to keep track of the collisions suffered by each particle, in the Langevin setup the picture is more coarse-grained: in each time-step $\Delta t$, the particle is given a random momentum kick, depending on the local value of the transport coefficients. However, it is possible to isolate the contribution to the anisotropy from the quarks decoupling at various values of the longitudinal proper-time $\tau\equiv\sqrt{t^2-z^2}$. The study is performed for Pb-Pb collisions at $\sqrt{s_{\rm NN}}\!=\!5.02$ TeV and Glauber-MC initial conditions, properly averaged depending on the considered flow-harmonic, as discussed in Sec.~\ref{sec:medium}.

In Fig.~\ref{fig:decoupling} we display the distribution of the decoupling time of the charm quarks in the 10-30\% centrality class. Notice (as can be seen from the time-integrated red curves) that half of the quarks escape from the fireball only after a quite long time $\tau\gsim 7$ fm/c and this holds for both choices of transport coefficients, which give rise to very similar curves. Only a small fraction of about 10\% of quarks spend in the medium a time $\lsim$ 4 fm/c. Hence, we expect that the interaction with the medium, in light of the average long time spent in the latter by the heavy quarks, provides a non negligible effect in determining the final angular distribution of their momenta.
  
%%%%%%%%%%%%%%%%%%%%%%%%%%
\begin{figure}[!ht]
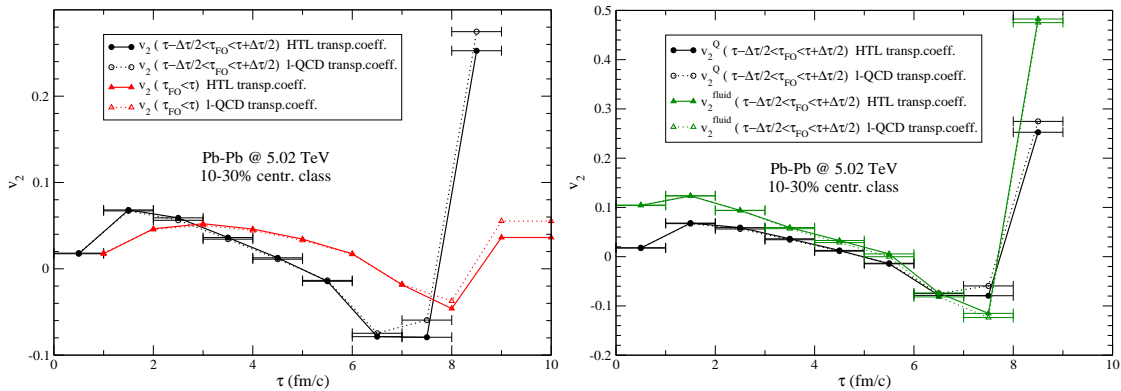

\begin{center}
\includegraphics[clip,width=0.48\textwidth]{v2-vs-FO_comp_10-30}
\includegraphics[clip,width=0.48\textwidth]{v2-Qvsfluid-FO_comp_10-30}
\caption{The elliptic flow of charm quarks decoupling at various values of the longitudinal proper-time $\tau$ in 10-30\% Pb-Pb collisions. The final integrated result (red curves with triangles in the left panel) comes from the interplay of opposite-sign contributions from the initial, intermediate and final stages of the fireball evolution. Differences between results obtained with weak-coupling and l-QCD transport coefficients look mild. In the right panel the heavy-quark $v_2$ is compared to the one of the fluid cells from which they decouple.}\label{fig:v2-Qfluid} 
\end{center}
\end{figure}
%%%%%%%%%%%%%%%%%%%%%%%%%%
\begin{figure}[!ht]
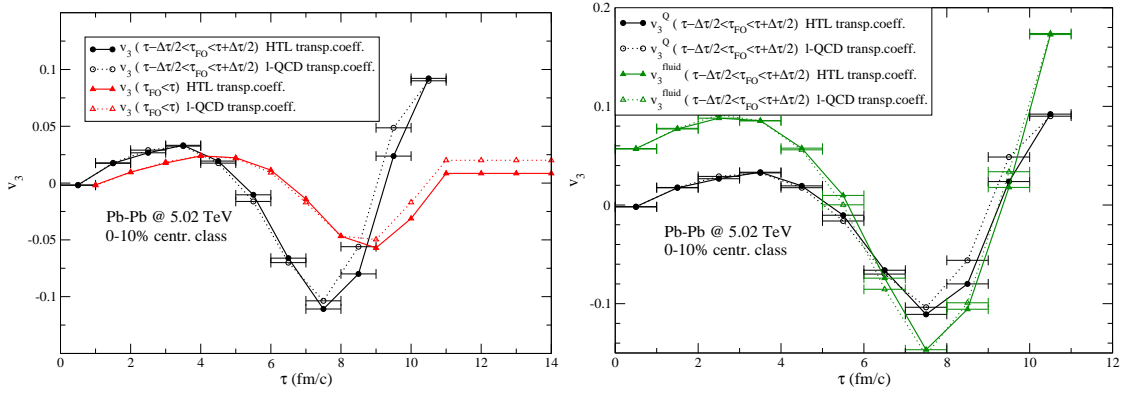

\begin{center}
\includegraphics[clip,width=0.48\textwidth]{v3-vs-FO_comp_0-10}
\includegraphics[clip,width=0.48\textwidth]{v3-Qvsfluid-FO_comp_0-10}
\caption{The triangular flow of charm quarks decoupling at various values of the longitudinal proper-time $\tau$ in 0-10\% Pb-Pb collisions. The final integrated result (red curves with triangles in the left panel) comes from the interplay of opposite-sign contributions from the initial, intermediate and final stages of the fireball evolution. Differences between results obtained with weak-coupling and l-QCD transport coefficients look mild. In the right panel the heavy-quark $v_3$ is compared to the one of the fluid cells from which they decouple. }\label{fig:v3-Qfluid} 
\end{center}
\end{figure}
%%%%%%%%%%%%%%%%%%%%%%%%%%
In Figs.~\ref{fig:v2-Qfluid} and~\ref{fig:v3-Qfluid} we display the differential contribution to the elliptic and triangular flow of charm quarks from particles decoupling at various values of the longitudinal proper-time $\tau$. In both cases the pattern is quite similar. Quarks decoupling very early ($\tau\lsim 2\!-\!3$ fm/c) provide a positive contribution, interpreted as arising from the previously discussed anisotropic escape probability. For larger values of the decoupling time the situation changes, and the Fourier coefficients start to decrease with increasing $\tau_{\rm FO}$, getting even negative until reaching a minimum around the time $\tau\approx 7$ fm/c at which most of the quarks decouple. One has then a sudden increase of the $v_2$ and $v_3$ of the heavy quarks decoupling during the latest stage, which makes the integrated final result positive. Interestingly, the picture depends only mildly on the transport coefficients (weak-coupling HTL or non-perturbative l-QCD) employed.
The peculiar behaviour of flow development can be interpreted also in light of the  freeze-out $\tau-r$ correlation plotted in the right panel of Fig.~\ref{fig:decoupling}, in which one can clearly identify two bands -- corresponding to heavy quarks decoupling along the $x$ and $y$-axis respectively -- which at a certain value of $\tau$ cross each other: at the very latest times only quarks moving along the $x$-axis have still to decouple and this gives rise to the very large contribution to the $v_2$ seen in the left panel of Fig.~\ref{fig:v2-Qfluid}. As it can be seen, in our framework in which one considers the heavy-quark propagation throughout a background medium undergoing an hydrodynamic expansion, the final flow signal (both for $v_2$ and $v_3$) is not dominated by the few particles escaping very early as in the study performed within a pure transport setup in~\cite{He:2015hfa}, but arises from the non-trivial interplay of opposite-sign contributions from all the different decoupling times. Interestingly, as can be seen from the green curves in the right panels of Figs.~\ref{fig:v2-Qfluid} and~\ref{fig:v3-Qfluid}, the trend of the $v_2$ and $v_3$ of the quarks looks in qualitative agreement with the collective elliptic and triangular flow ($v_{2/3}^{\rm fluid}$) of the fluid cells from which they decouple.

%%%%%%%%%%%%%%%%%%%%%%%%%%
\begin{figure}[!ht]
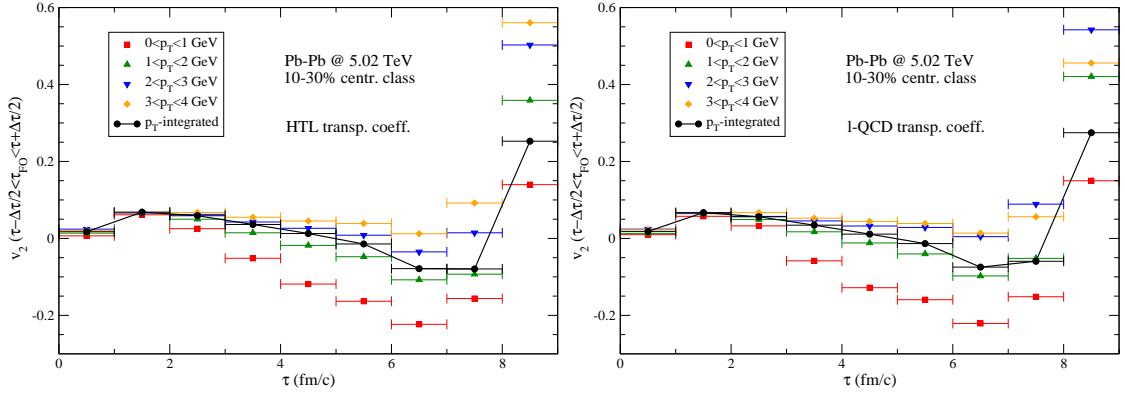

\begin{center}
\includegraphics[clip,width=0.48\textwidth]{v2-vs-FOdiff-pTdiff-HTL}
\includegraphics[clip,width=0.48\textwidth]{v2-vs-FOdiff-pTdiff-LAT}
\caption{The elliptic flow of charm quarks decoupling at various values of the longitudinal proper-time $\tau$ in 10-30\% semi-central Pb-Pb collisions for different $p_T$-bins with HTL (left panel) and l-QCD (right panel) transport coefficients. Differences arise at late time and for the hardest particles, due to the different momentum-dependence of the transport coefficients.}\label{fig:v2_taudiff} 
\end{center}
\end{figure}
%%%%%%%%%%%%%%%%%%%%%%%%%%
\begin{figure}[!ht]
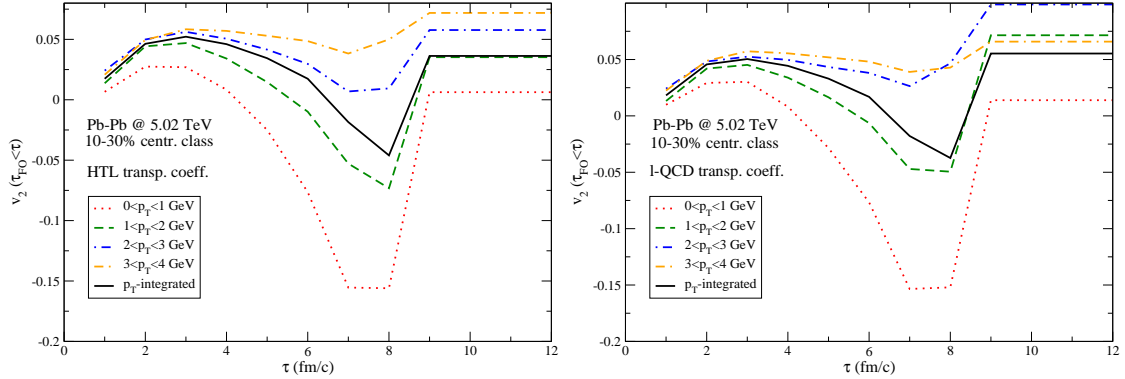

\begin{center}
\includegraphics[clip,width=0.48\textwidth]{v2-vs-FO-pTdiff-HTL}
\includegraphics[clip,width=0.48\textwidth]{v2-vs-FO-pTdiff-LAT}
\caption{Time development of the elliptic flow in 10-30\% semi-central Pb-Pb collisions arising from the sum of the contribution of all charm quarks having decoupled from $\tau_0$ up to time $\tau$. Results obtained with HTL (left panel) and l-QCD (right panel) transport coefficients are displayed.}\label{fig:v2_tauint} 
\end{center}
\end{figure}
In Figs.~\ref{fig:v2_taudiff} and~\ref{fig:v2_tauint} we show the findings of a $p_T$-differential study of the time-development of the elliptic flow, considering both the time-differential and integrated results, respectively. As can be seen, at early times, when the signal is dominated by the anisotropic escape probability of the partons, the $p_T$-dependence of the effect is negligible, whereas it gets important during the later stages, where the interaction with the medium affects differently the propagation of heavy quarks of different momenta. Similar considerations hold for the dependence on the transport coefficients, HTL and l-QCD results being significantly different only at late times.

%%%%%%%%%%%%%%%%%%%%%%%%%%
\section{Model results versus experimental data}\label{sec:results}
Here we display some new results obtained with our setup, pushing its prediction from RHIC ($\sqrt{s_{\rm NN}}\!=\!200$ GeV) to top LHC energies ($\sqrt{s_{\rm NN}}\!=\!5.02$ TeV) and focusing mainly on the elliptic and (for the first time) triangular flow of charm quarks and hadrons.

\begin{figure}[!ht]
\begin{center}
\includegraphics[clip,height=6cm]{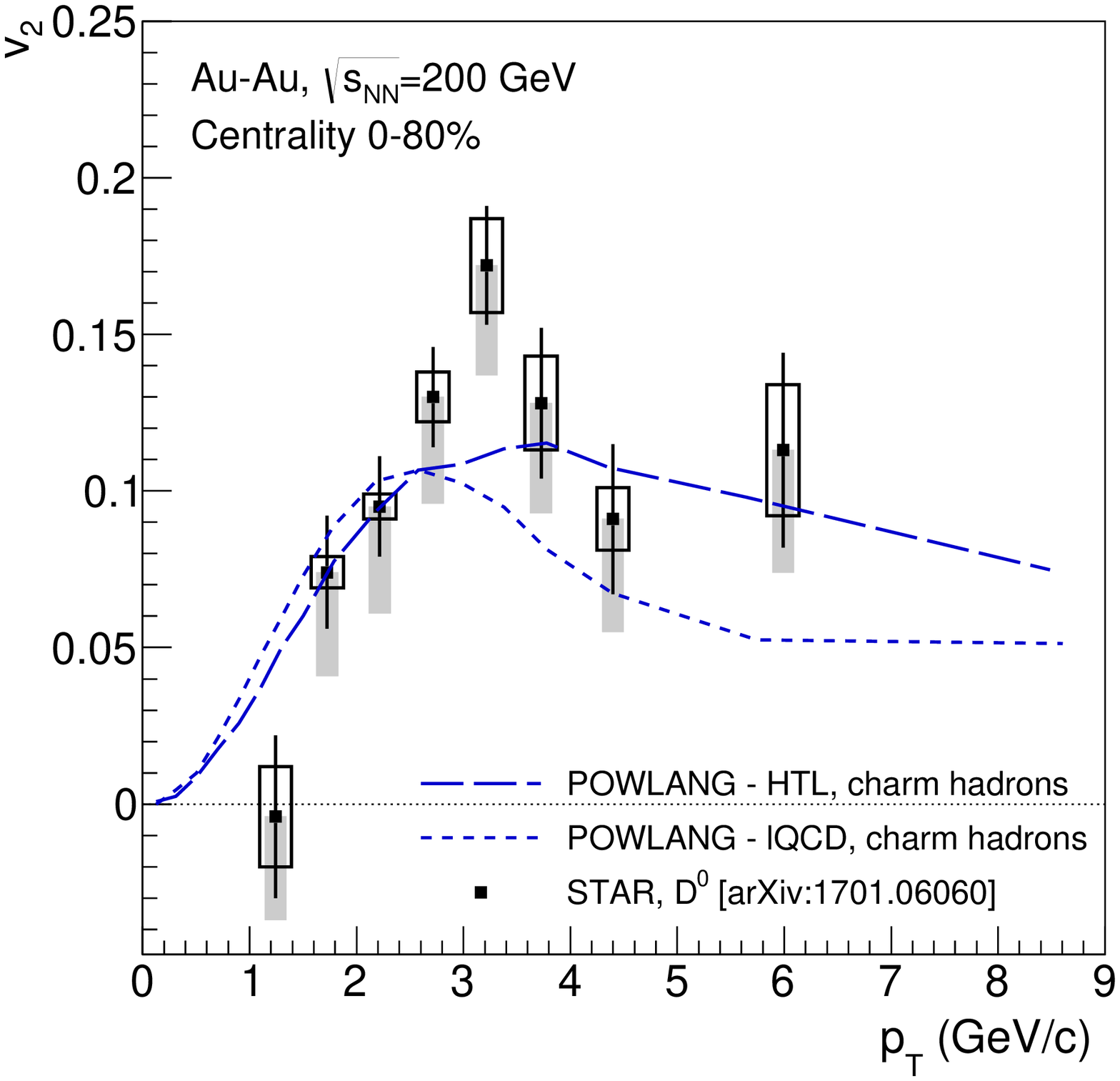}
\includegraphics[clip,height=6cm]{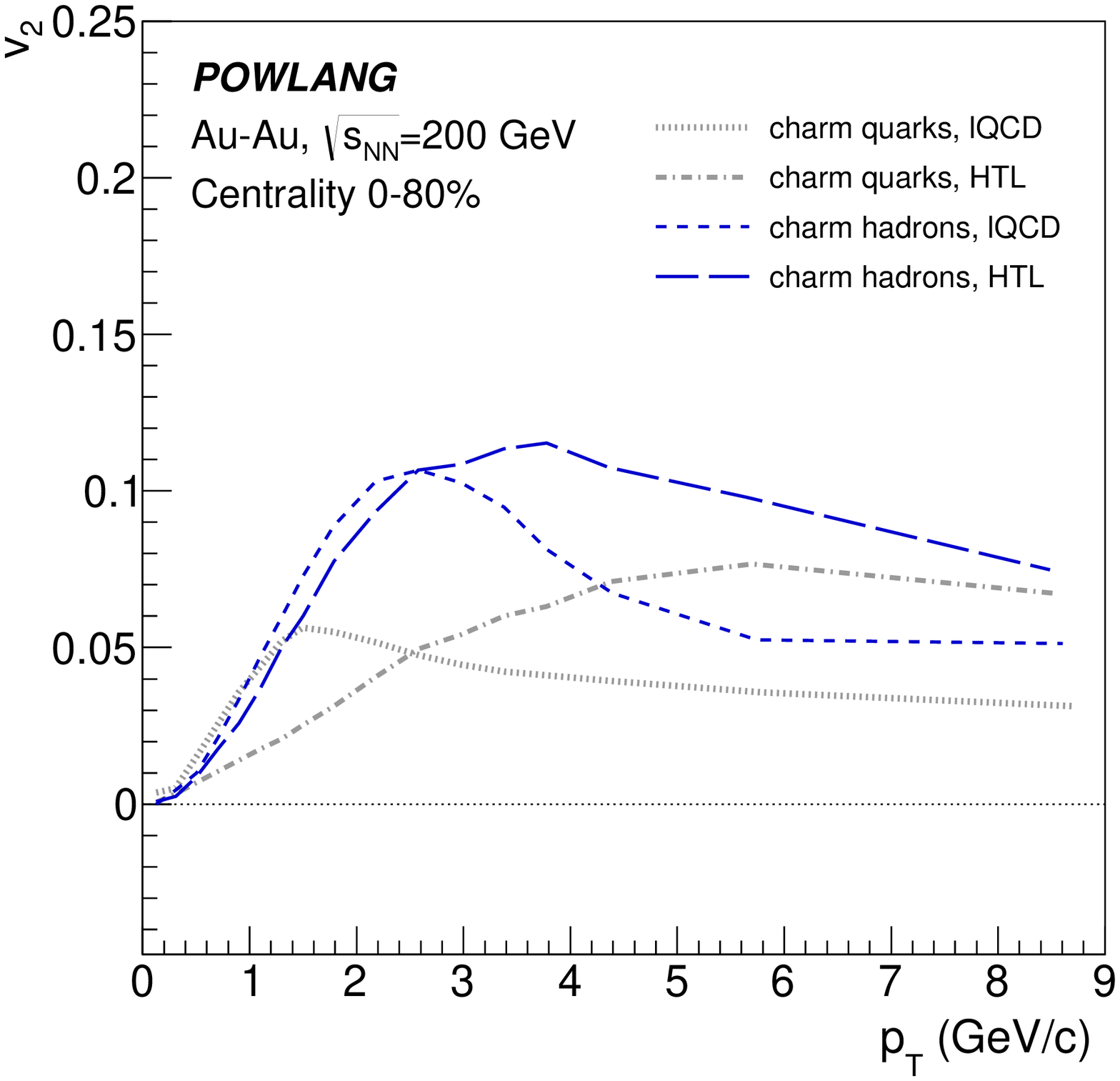}
\caption{The elliptic flow of $D^0$ mesons in 0-80\% Au-Au collisions at  $\sqrt{s_{\rm NN}}\!=\!200$ GeV. Our results with HTL and l-QCD transport coefficients are compared to recent STAR data~\cite{Adamczyk:2017xur}. The right panel illustrates the effect of in-medium hadronization, which enhances the anisotropy due to the additional flow inherited from the light thermal partons.}\label{fig:v2STAR} 
\end{center}
\end{figure}
In Fig.~\ref{fig:v2STAR} we display the POWLANG predictions for the elliptic flow of charm quarks and hadrons in Au-Au collisions at $\sqrt{s_{\rm NN}}\!=\!200$ GeV in the 0-80\% centrality class, comparing our results to STAR measurements of the $D^0$ meson $v_2$~\cite{Adamczyk:2017xur}. In a previous publication~\cite{Beraudo:2014boa} we already showed how the flow inherited from the light thermal partons at hadronization via in-medium recombination was able to boost the spectra of charmed hadrons towards slightly larger values of $p_T$, leading to an enhancement of the $D^0$ $R_{\rm AA}$ at intermediate $p_T$ not observed in the results obtained with independent vacuum fragmentation functions. As can be seen in Fig.~\ref{fig:v2STAR}, a similar effect occurs for the elliptic flow. In POWLANG the elliptic flow of charm quarks at the end of the partonic phase is non-negligible, but not sufficient to describe the sizable $D^0$ $v_2$ measured in the experiment. Notice that, at the quark level, results obtained with weak-coupling HTL and non-perturbative l-QCD transport coefficients differ substantially, the l-QCD curve displaying a much larger $v_2$ at low $p_T$ due to the larger value of the momentum diffusion coefficient (this can be also appreciated comparing the left and right panels of Fig.~\ref{fig:v2_tauint}), the HTL curve saturating instead at a larger value of $v_2$ at high $p_T$, simply reflecting the different amount of parton energy-loss in-plane versus out-of-plane, larger in the HTL case due to the steep rise of $\kappa_\|(p)$. On the other hand, after hadronization via recombination with light thermal partons feeling the collective expansion of the medium, the $v_2$ of charmed hadrons turns out to increase at low-moderate $p_T$ and looks in better agreement with the experimental data.

%%%%%%%%%%%%%%%%%%%%%%%%%%%%%%%%%%%%%%%
\begin{figure}[!ht]
\begin{center}
\includegraphics[clip,height=4.8cm]{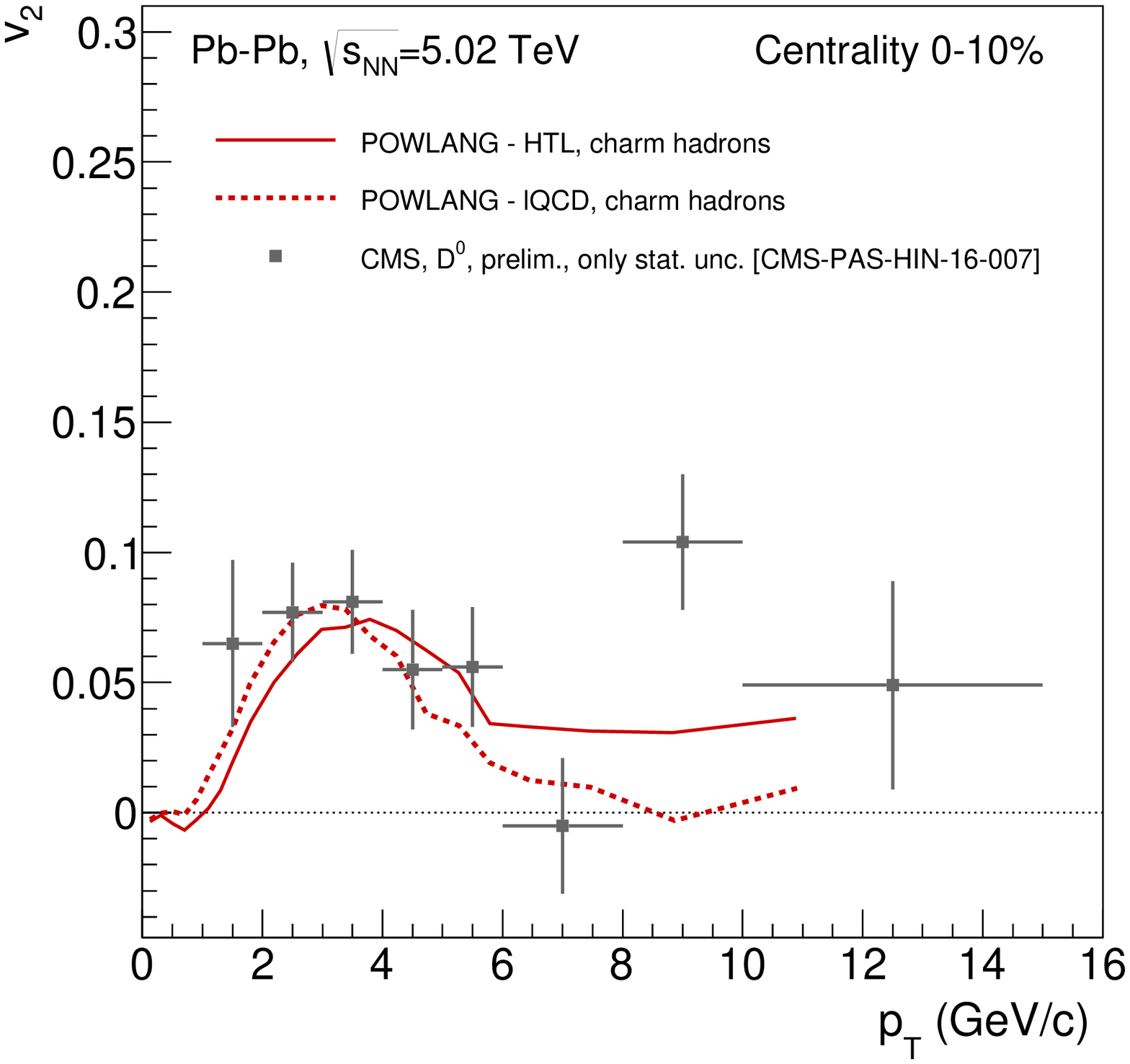}
\includegraphics[clip,height=4.8cm]{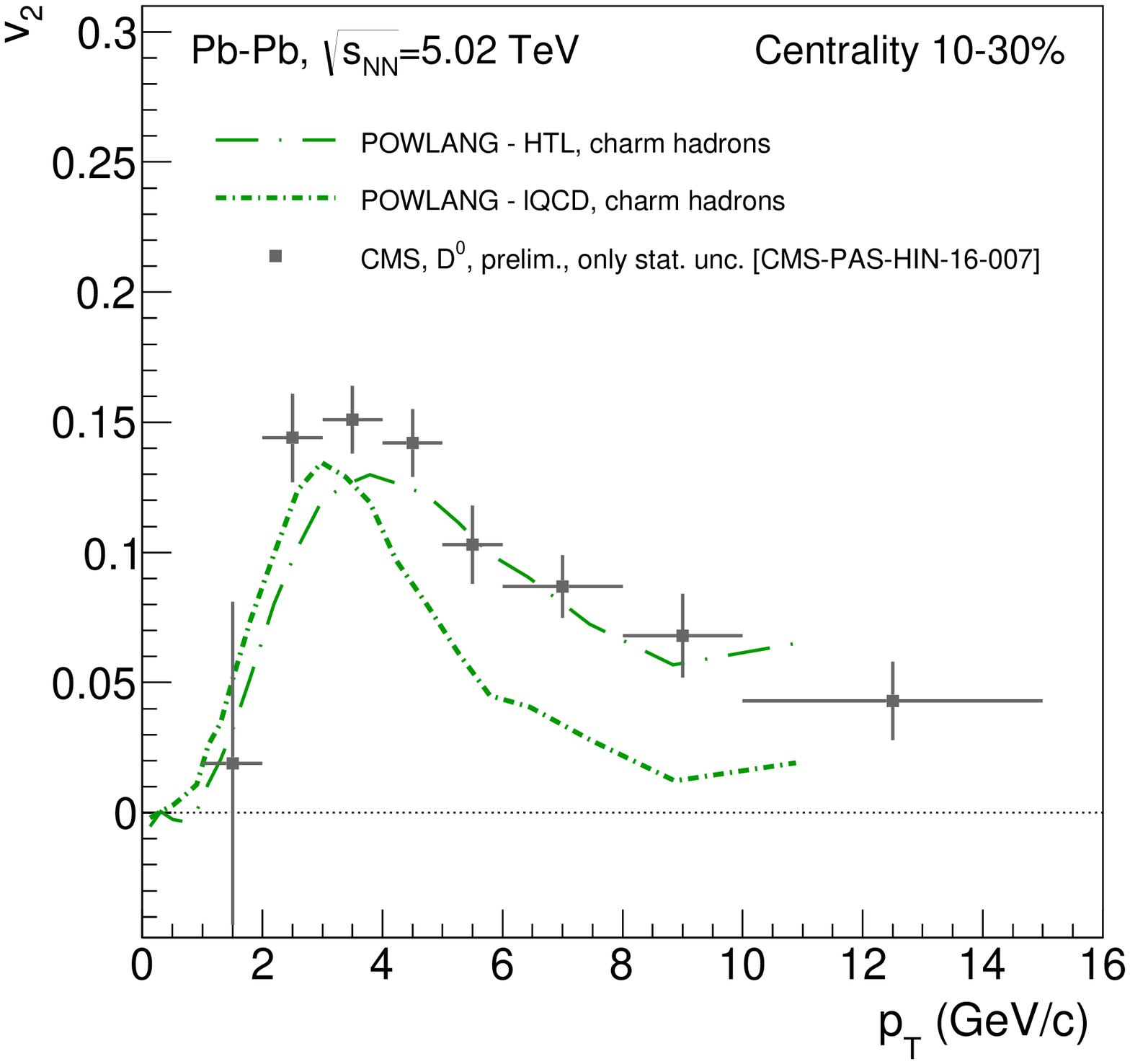}
\includegraphics[clip,height=4.8cm]{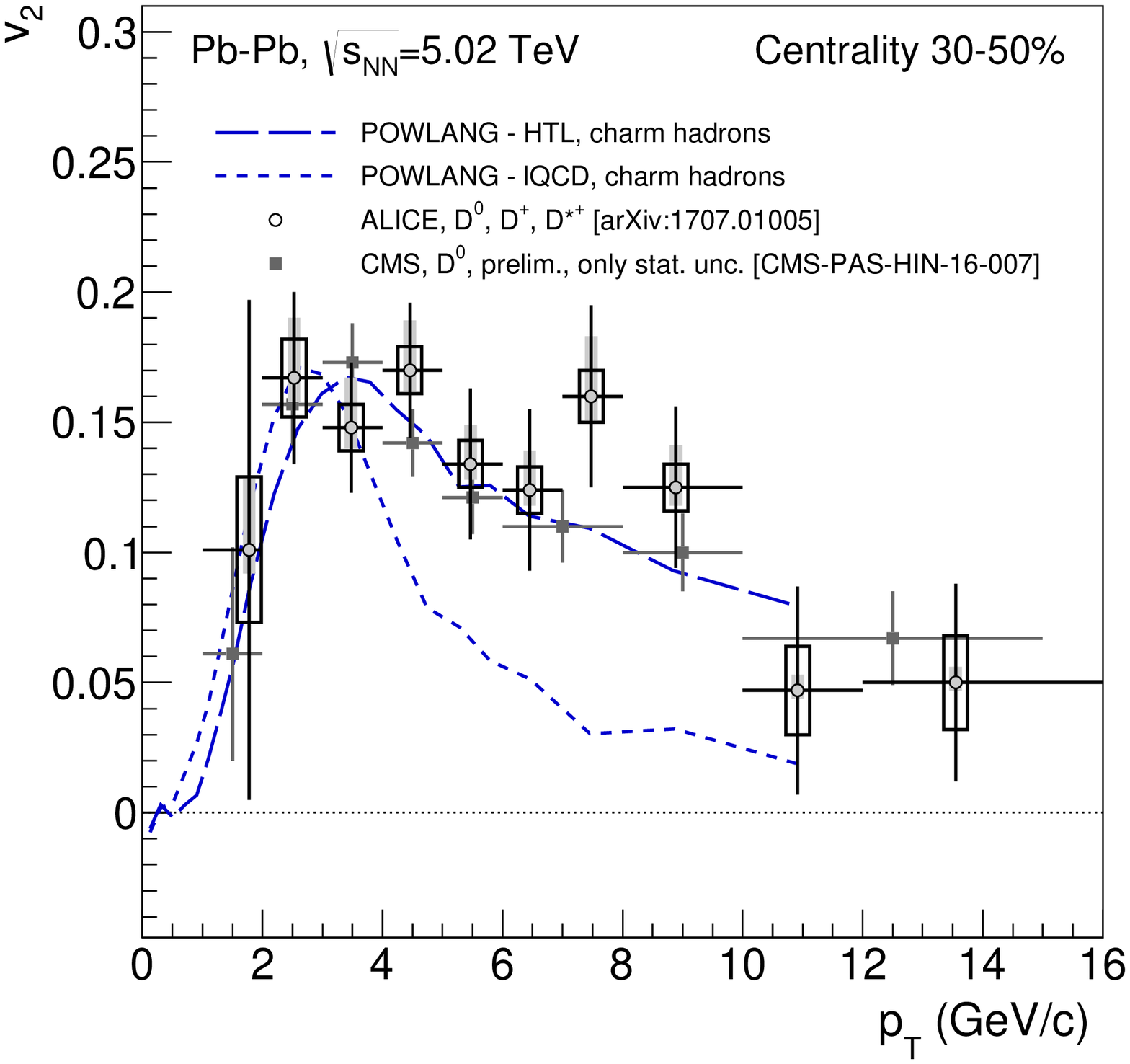}
\caption{The elliptic flow of $D^0$ mesons in Pb-Pb collisions at $\sqrt{s_{\rm NN}}\!=\!5.02$ TeV, for various centrality classes. POWLANG predictions with HTL and l-QCD transport coefficients are compared to ALICE~\cite{Acharya:2017qps} and CMS data~\cite{Sirunyan:2017plt}.}\label{fig:v2vsCMS&ALICE} 
\end{center}
\end{figure}
%%%%%%%%%%%%%%%%%%%%%%%%%%%%%%%%%%%%%%
\begin{figure}[!ht]
\begin{center}
\includegraphics[clip,height=4.8cm]{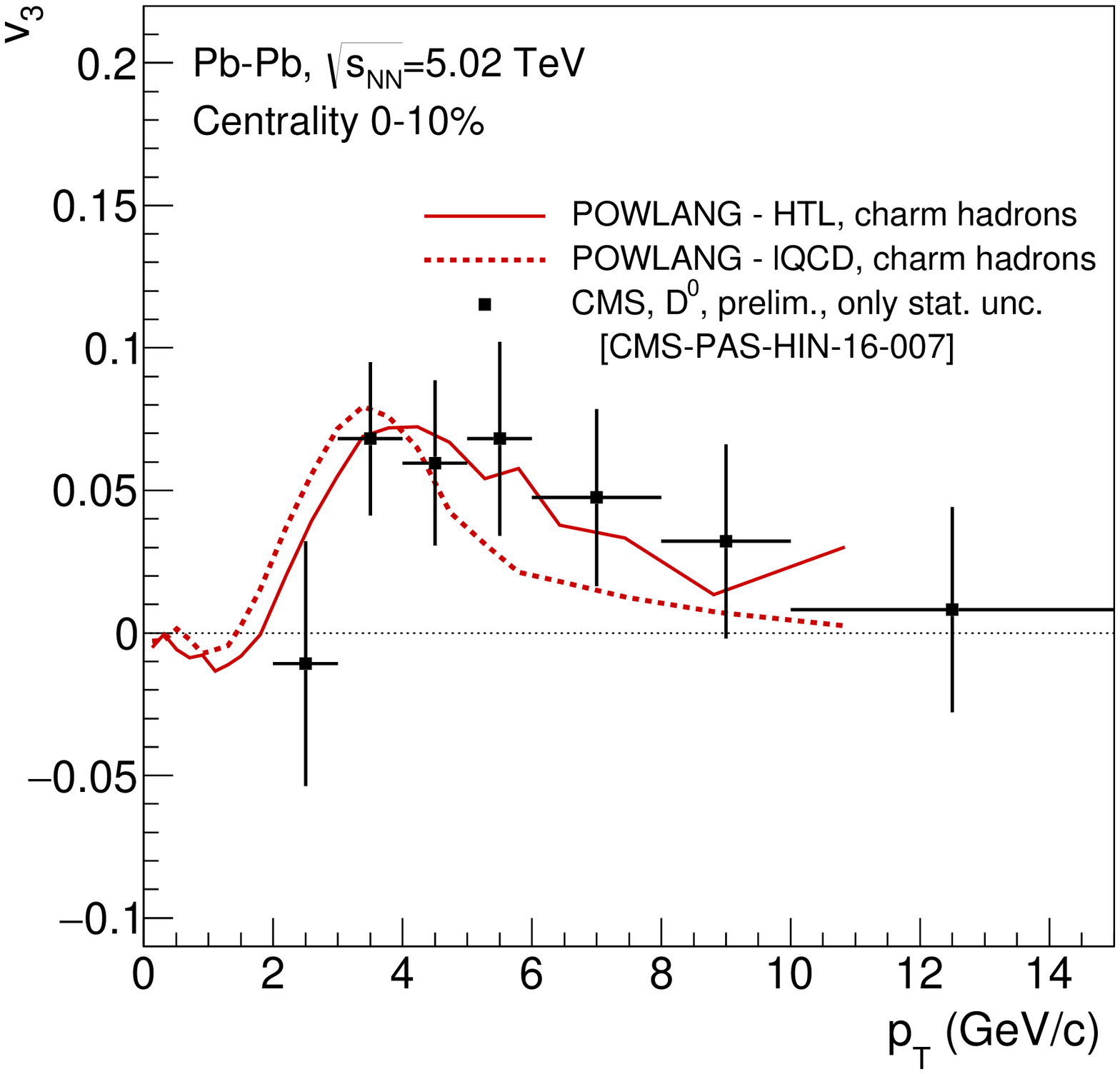}
\includegraphics[clip,height=4.8cm]{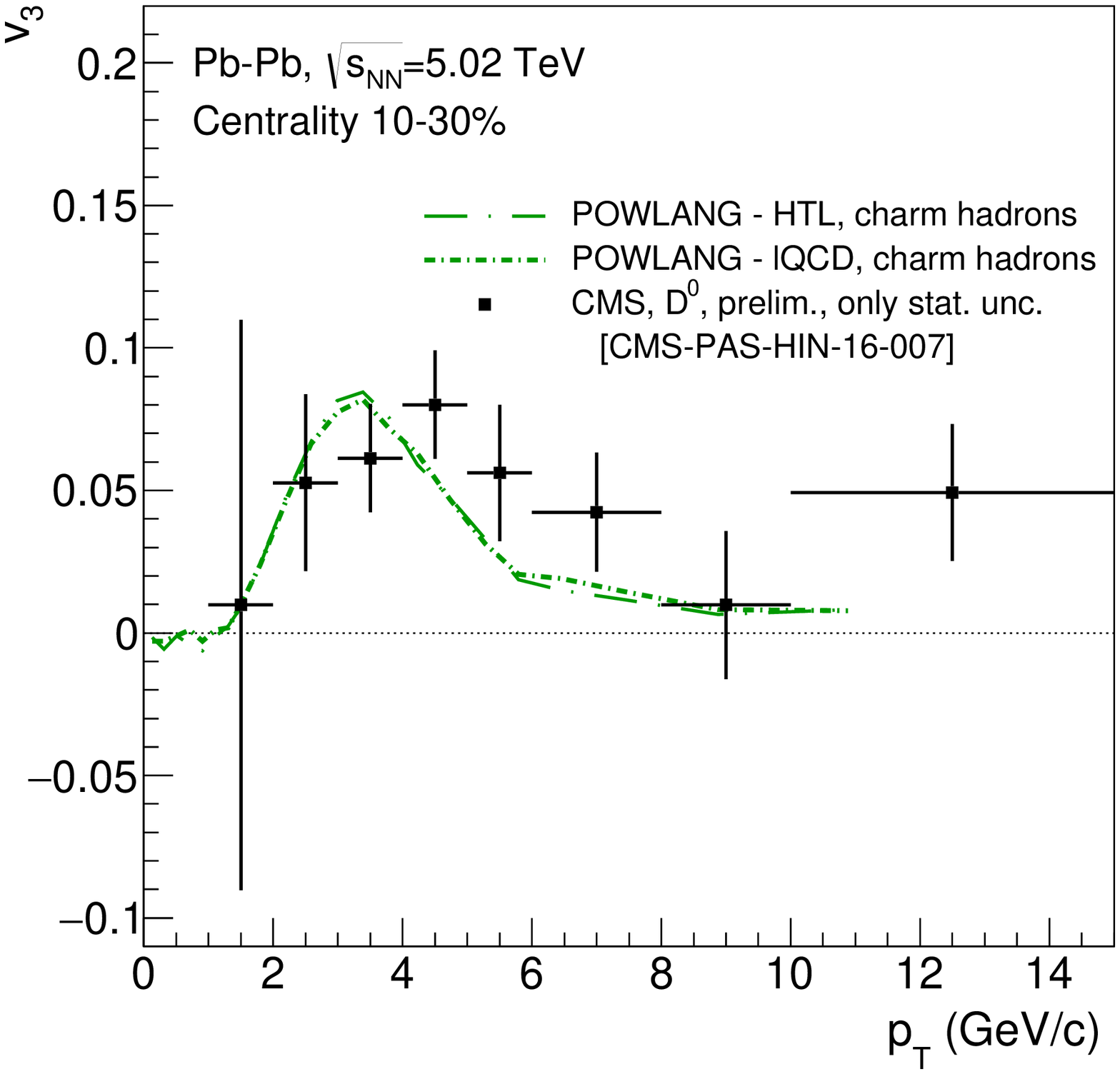}
\includegraphics[clip,height=4.8cm]{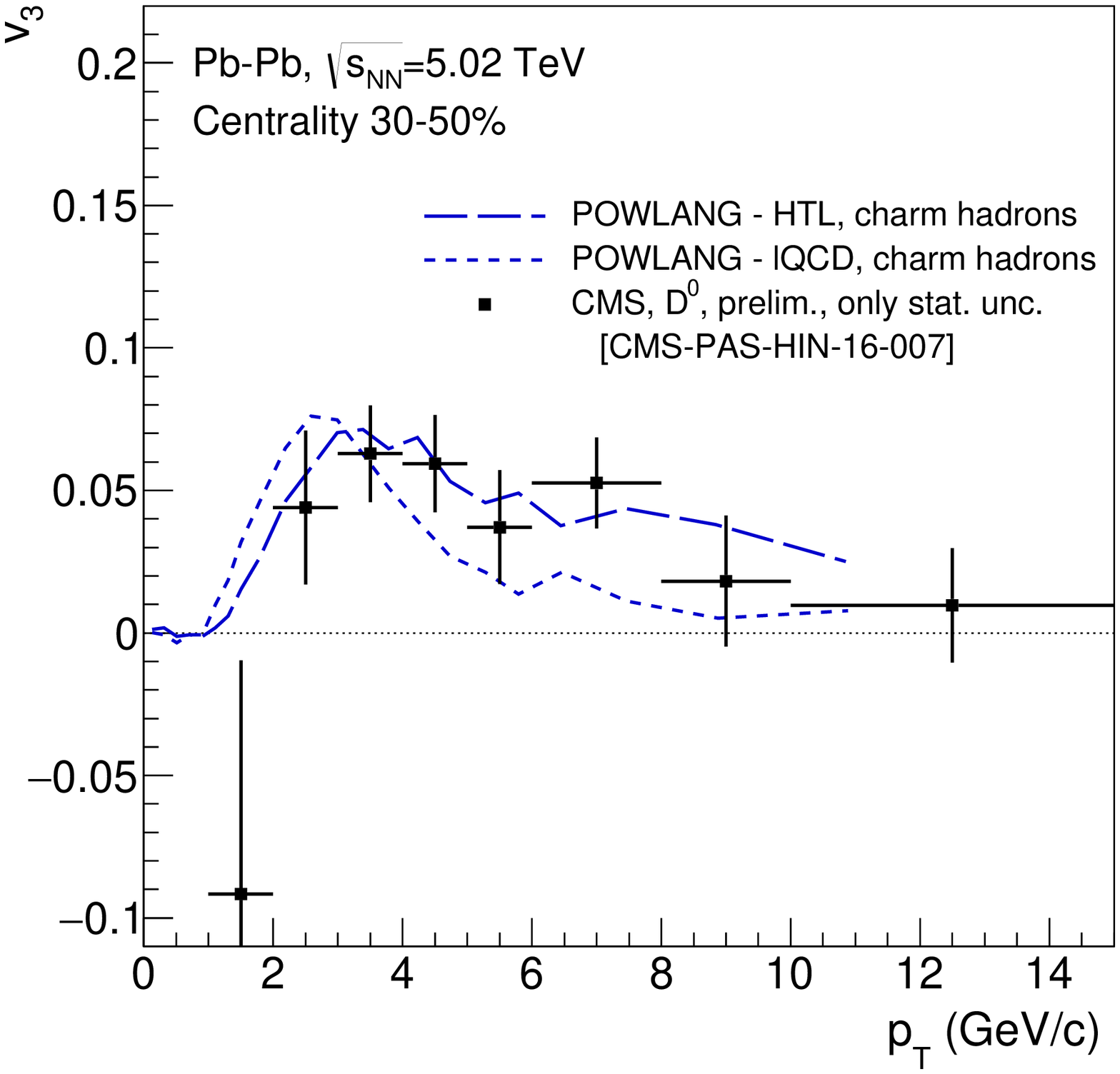}
\caption{The triangular flow of $D^0$ mesons in Pb-Pb collisions at $\sqrt{s_{\rm NN}}\!=\!5.02$ TeV, for various centrality classes. The signal arises from event-by-event fluctuations in the initial conditions. POWLANG predictions with HTL and l-QCD transport coefficients are compared to CMS data~\cite{Sirunyan:2017plt}.}\label{fig:v3vsCMS} 
\end{center}
\end{figure}

In Figs.~\ref{fig:v2vsCMS&ALICE} and \ref{fig:v3vsCMS} we consider POWLANG predictions for the elliptic and triangular flow of charmed hadrons in Pb-Pb collisions at $\sqrt{s_{\rm NN}}\!=\!5.02$ TeV. As in the previous case, differences between HTL and l-QCD results are more evident at high $p_T$, due to the different momentum dependence of the transport coefficients, in particular of $\kappa_\|$. For the background medium we employ Glauber-MC initial conditions, taking, as explained in detail in Sec.~\ref{sec:medium}, a proper weighted average of hundreds of collisions belonging to the same centrality class. The agreement with the $D$-meson $v_2$ and $v_3$ values measured by the ALICE~\cite{Acharya:2017qps} and CMS~\cite{Sirunyan:2017plt} collaborations is quite good. As in the case of light hadrons, the triangular flow does not arise from the finite impact parameter of the collisions (in fact the signal does not change so much in the different centrality classes) but is due to event-by-event fluctuations: in our study we limited ourselves to geometric fluctuations in the nucleon positions.

%%%%%%%%%%%%%%%%%%%%%%%%%%%%%%%%%%%%%%
\begin{figure}[!ht]
\begin{center}
\includegraphics[clip,height=6cm]{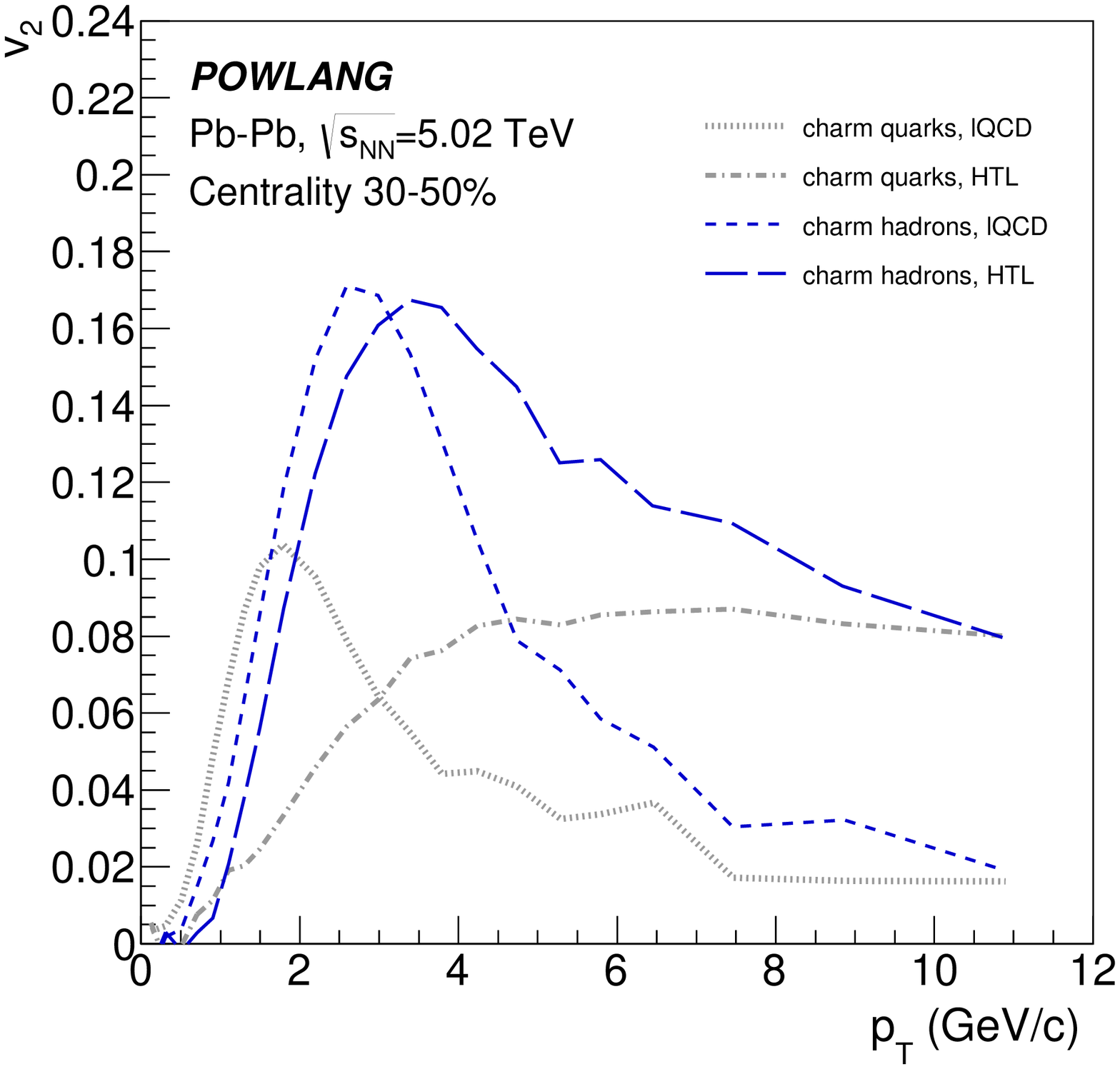}
\includegraphics[clip,height=6cm]{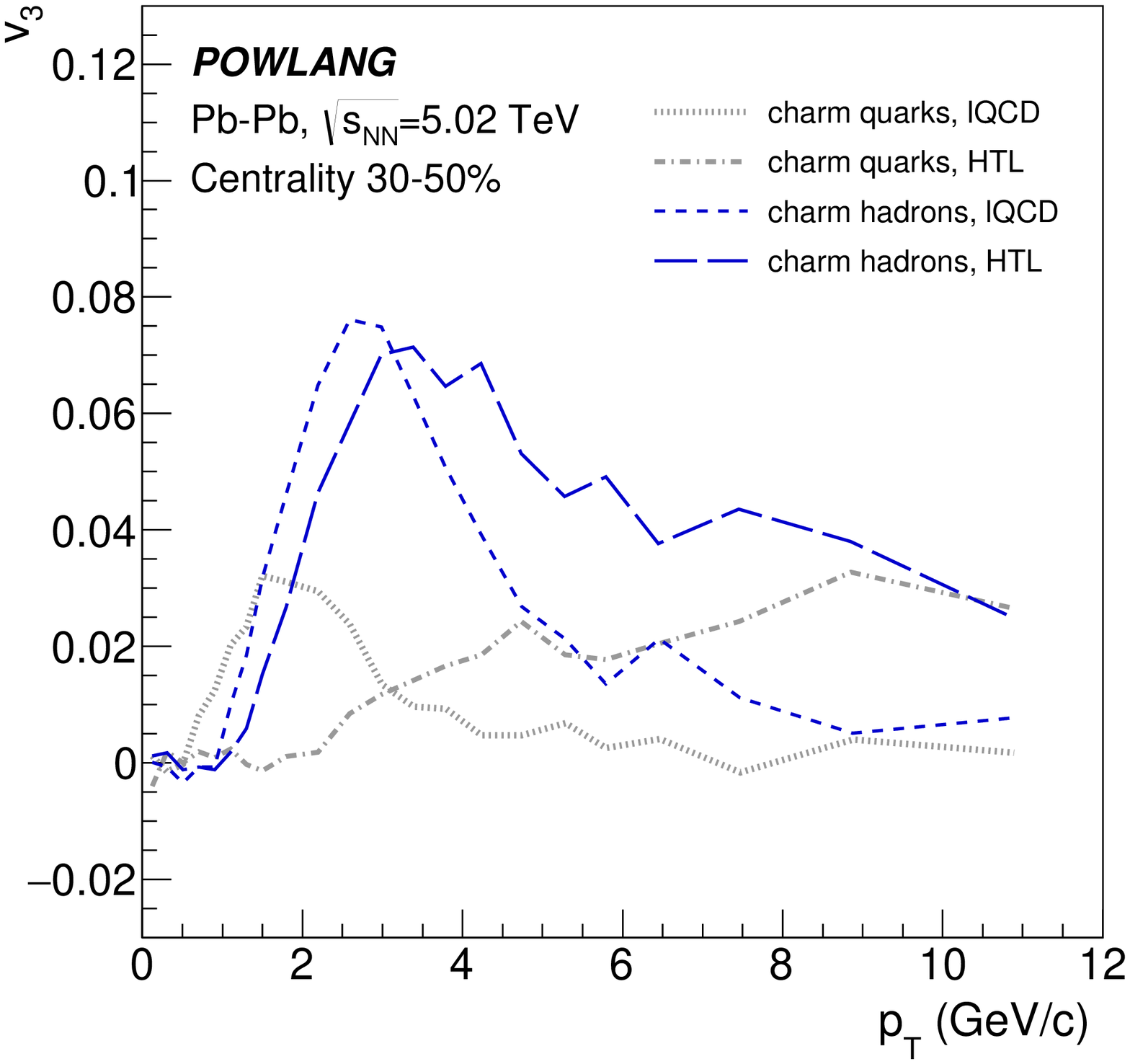}
\caption{The elliptic (left panel) and triangular (right panel) flow of charm quarks and hadrons in Pb-Pb collisions at  $\sqrt{s_{\rm NN}}\!=\!5.02$ TeV. The additional flow inherited at hadronization through recombination with light partons from the medium provides a significant contribution to the final signal.}\label{fig:flow_cD} 
\end{center}
\end{figure}
Also in this case we can disentangle in the model the effect of the heavy-quark transport through the deconfined plasma and of the in-medium hadronization, both for the elliptic and the triangular flow; the results are shown in Fig.~\ref{fig:flow_cD}. Similar considerations to what found at $\sqrt{s_{\rm NN}}\!=\!200$ GeV hold. The additional flow acquired at hadronization via recombination with light thermal partons enhances the $D$-meson anisotropies at low and intermediate $p_T$, moving the curves closer to the experimental data. Also in this case, at the partonic level, the flow signal at low $p_T$ obtained with l-QCD transport coefficients is larger due to the stronger coupling with the medium; notice however that in this kinematic range hadronization tends to wash out the differences arising from the transport coefficients. On the other hand, at high $p_T$ larger $v_2$ and $v_3$ values are obtained in the HTL case, reflecting the larger energy-loss at high momentum.

%%%%%%%%%%%%%%%%%%%%%%%%%%%%%%%%%%%%%%%%%%%%%%%
\begin{figure}[!ht]
\begin{center}
\includegraphics[clip,height=6cm]{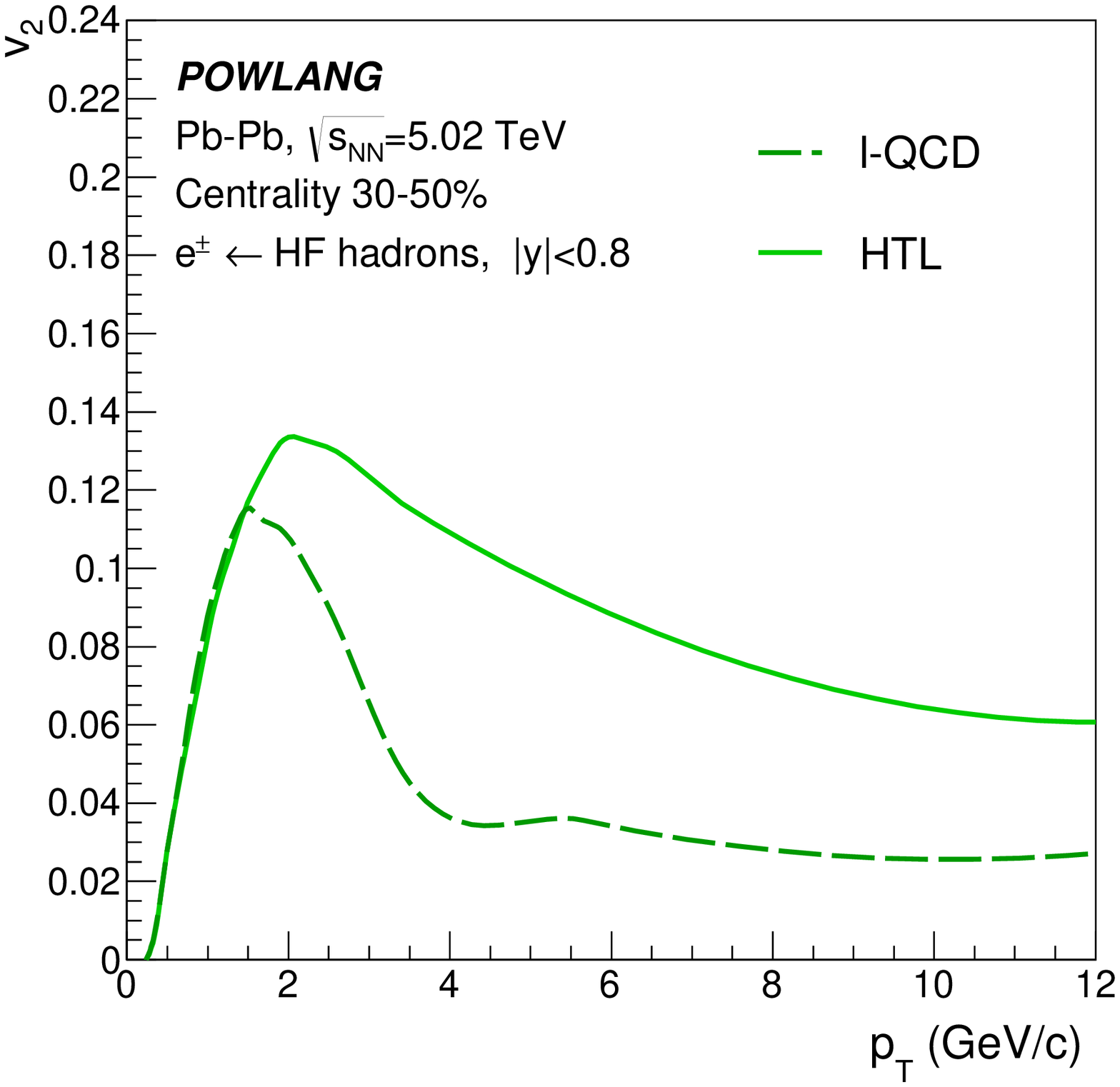}
\includegraphics[clip,height=6cm]{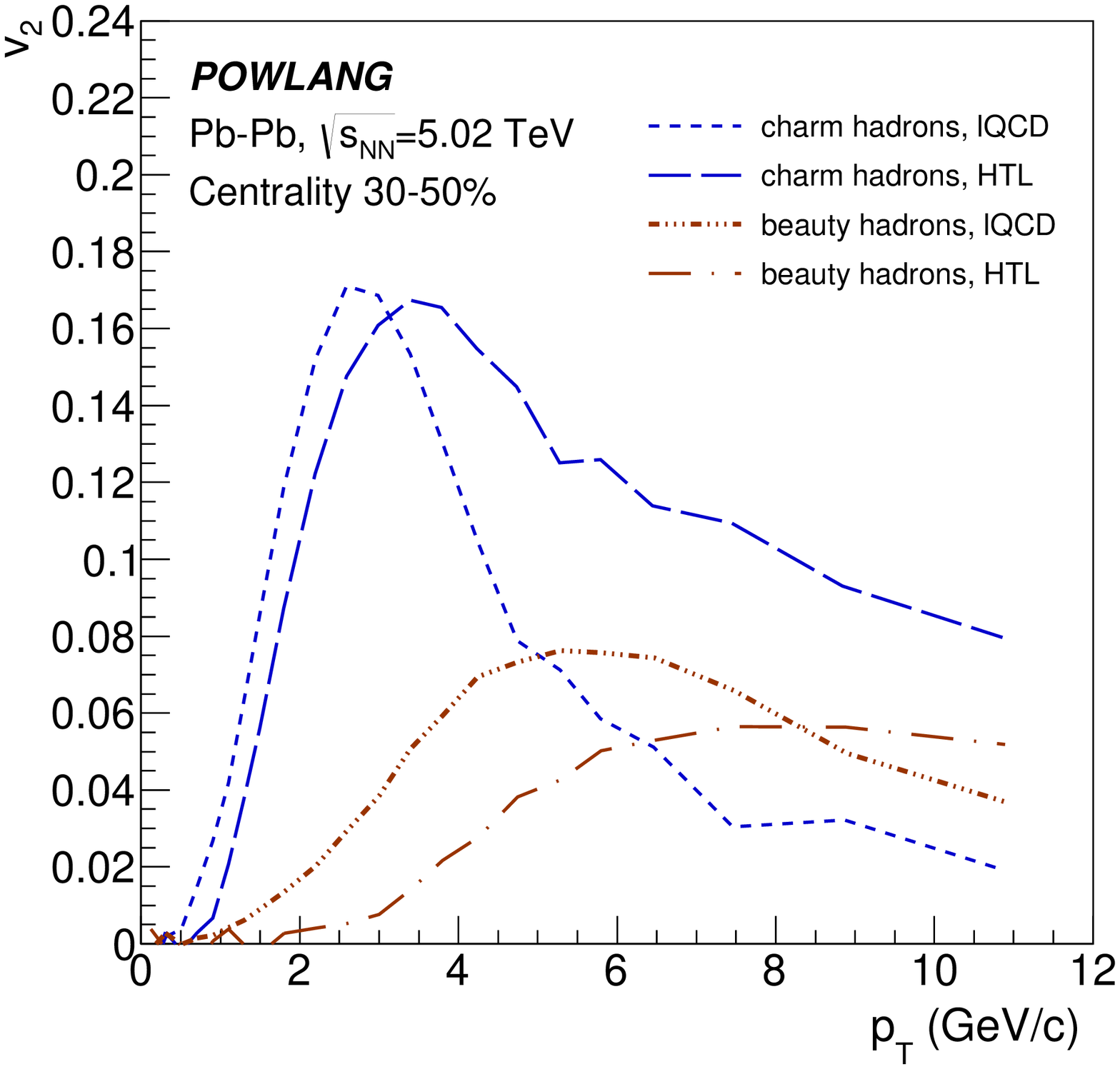}
\caption{The elliptic flow of electrons from semi-leptonic decays of heavy-flavour hadrons in non-central Pb-Pb collisions at $\sqrt{s_{\rm NN}}\!=\!5.02$ TeV. POWLANG results with HTL and l-QCD transport coefficients are compared. In the right panel we display the corresponding results for the flow of the parent charm and beauty hadrons.}\label{fig:flow_HFE} 
\end{center}
\end{figure}
In Fig.~\ref{fig:flow_HFE} we address the elliptic flow of electrons from heavy-flavour semi-leptonic decays in the case of non-central Pb-Pb collisions at $\sqrt{s_{\rm NN}}\!=\!5.02$ TeV. Curves with HTL and lattice-QCD transport coefficients look very similar at low momentum, while they display a quite different behaviour at higher $p_T$ arising mainly from the one of the parent charm hadrons (see right panel).

%%%%%%%%%%%%%%%%%%%%%%%%%%%%%%%%%%%%%%%%%%%%%%%
\begin{figure}[!ht]
\begin{center}
\includegraphics[clip,height=6cm]{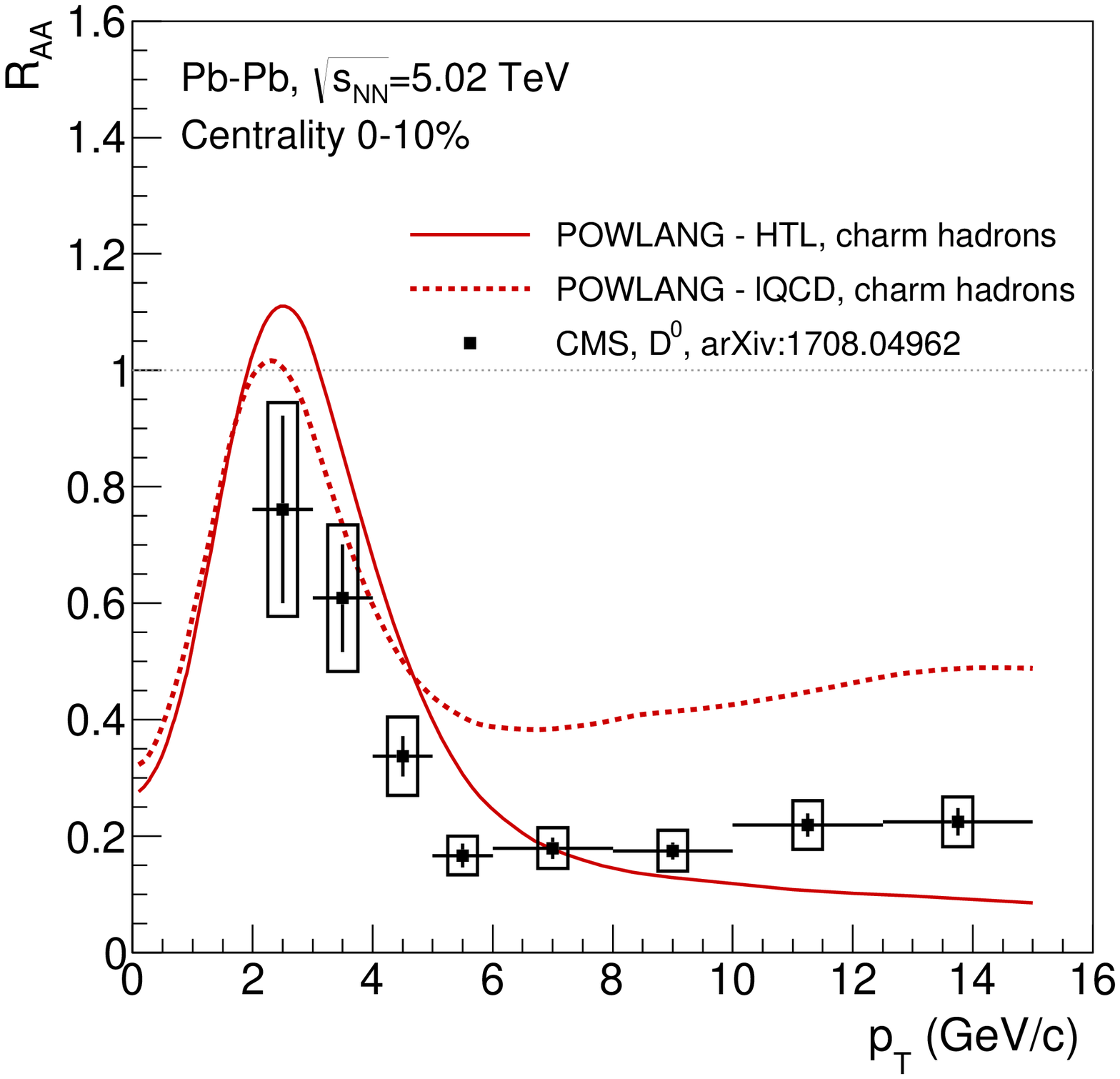}
\includegraphics[clip,height=6cm]{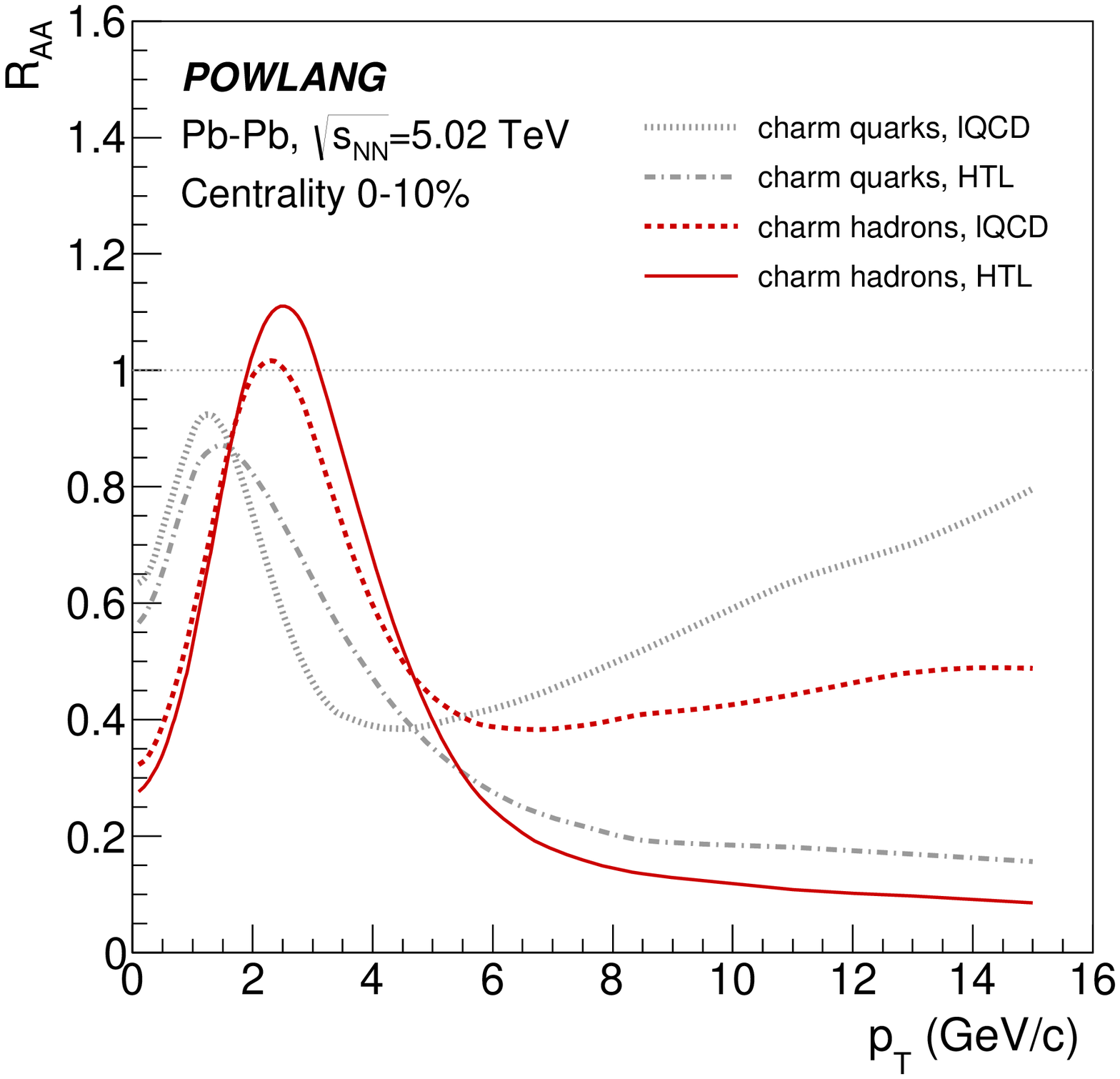}
\caption{POWLANG results with weak-coupling and l-QCD transport coefficients for the nuclear modification factor of charm hadron (left panel) and quark (right panel) spectra in central Pb-Pb collisions at $\sqrt{s_{\rm NN}}\!=\!5.02$ TeV. Transport-model predictions are compared to CMS data for $D^0$-mesons~\cite{Sirunyan:2017xss}.}\label{fig:RAA-vs-CMS} 
\end{center}
\end{figure}
%%%%%%%%%%%%%%%%%%%%%%%%%%%%%%%%%%%%%%%%%%%%%%%%%
\begin{figure}[!ht]
\begin{center}
\includegraphics[clip,height=6cm]{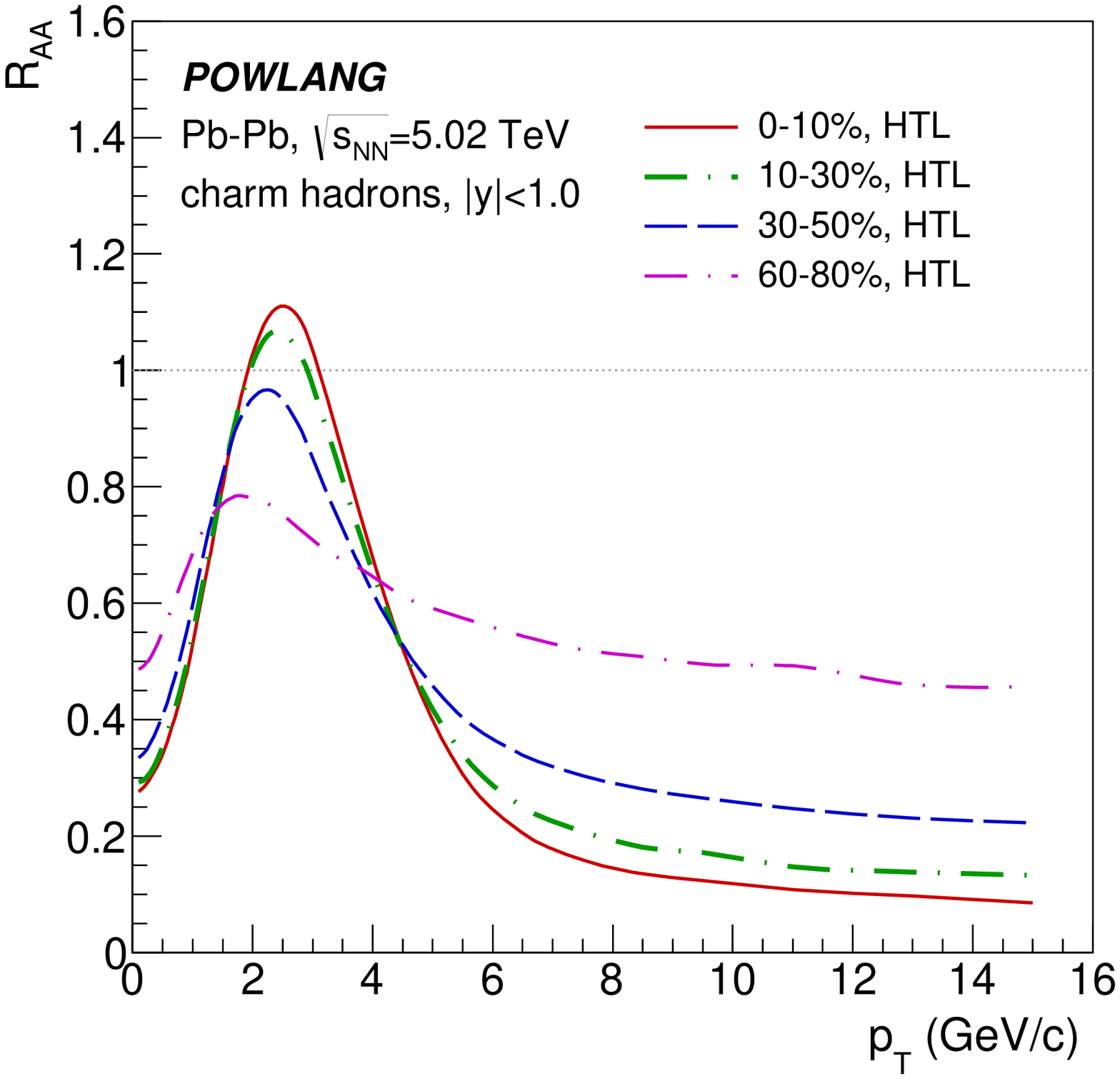}
\includegraphics[clip,height=6cm]{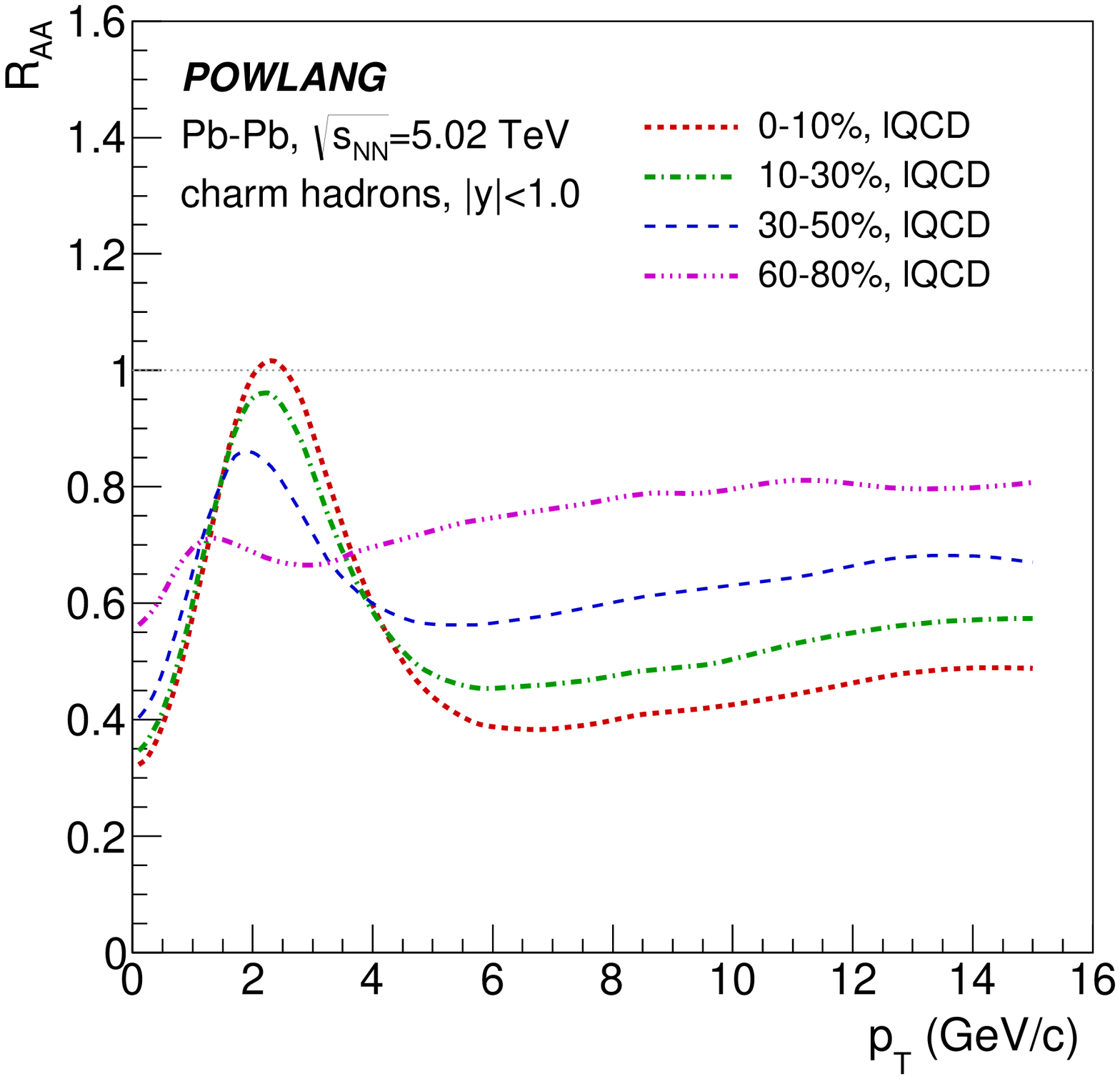}
\caption{POWLANG results with weak-coupling (left panel) and l-QCD (right panel) transport coefficients for the nuclear modification factor of charm hadron spectra in Pb-Pb collisions at $\sqrt{s_{\rm NN}}\!=\!5.02$ TeV for various centrality classes.}\label{fig:RAA-vs-centrality} 
\end{center}
\end{figure}
%%%%%%%%%%%%%%%%%%%%%%%%%%%%%%%%%%%%%%%%%%%%%%%
As usual, it is important to check that the same transport setup provides a consistent description not only of the azimuthal anisotropies of heavy-flavour hadron distributions, but also of the medium modifications of their $p_T$-spectra, reflecting, depending on the kinematic region, either the radial flow (dominant at low-moderate $p_T$) or the energy loss (the relevant effect at high-$p_T$) acquired/suffered by the heavy particles. Hence, in Figs.~\ref{fig:RAA-vs-CMS} and~\ref{fig:RAA-vs-centrality} we display the POWLANG predictions for the $R_{\rm AA}$ of charmed hadrons (and parent quarks) in Pb-Pb collisions at $\sqrt{s_{\rm NN}}\!=\!5.02$ TeV for various centrality classes, from central to peripheral ones. In the 0-10\% centrality class our transport results are compared to experimental measurements of the nuclear modification factor of $D^0$ mesons performed by the CMS collaboration~\cite{Sirunyan:2017xss}. Transport results are characterized by a pronounced peak (supported also by the available experimental data) around $p_T\!\approx\!3$ GeV/c, which we interpret as due to the radial flow, acquired in part crossing the deconfined medium (whose collective motion tend to boost the heavy quarks), in part at hadronization. This looks evident from the right panel of Fig.~\ref{fig:RAA-vs-CMS}, in which the bump in the charm hadron $R_{\rm AA}$ looks shifted to larger $p_T$ with respect to the corresponding partonic one. In POWLANG, as already discussed, hadronization is modeled through the formation of color-singlet strings/clusters obtained via recombination of the heavy quarks with light thermal partons flowing with the medium: hence, this provides a further boost to spectrum, which causes the bump to move to larger $p_T$. Transport results at high momenta display a strong sensitity to the choice of the transport coefficients. Weak-coupling HTL results, due to the steep rise of the longitudinal momentum-broadening, tend to overpredict the amount of energy-loss. On the other hand, information on the momentum dependence of the non-perturbative lattice-QCD result for $\kappa$ is missing and keeping it as a constant leads to a too small friction force acting on the heavy quarks at high momentum and hence to underestimate the energy loss. Experimental data suggest that reality sits perhaps in between these two scenarios. 

\begin{figure}[!ht]
\begin{center}
\includegraphics[clip,height=6cm]{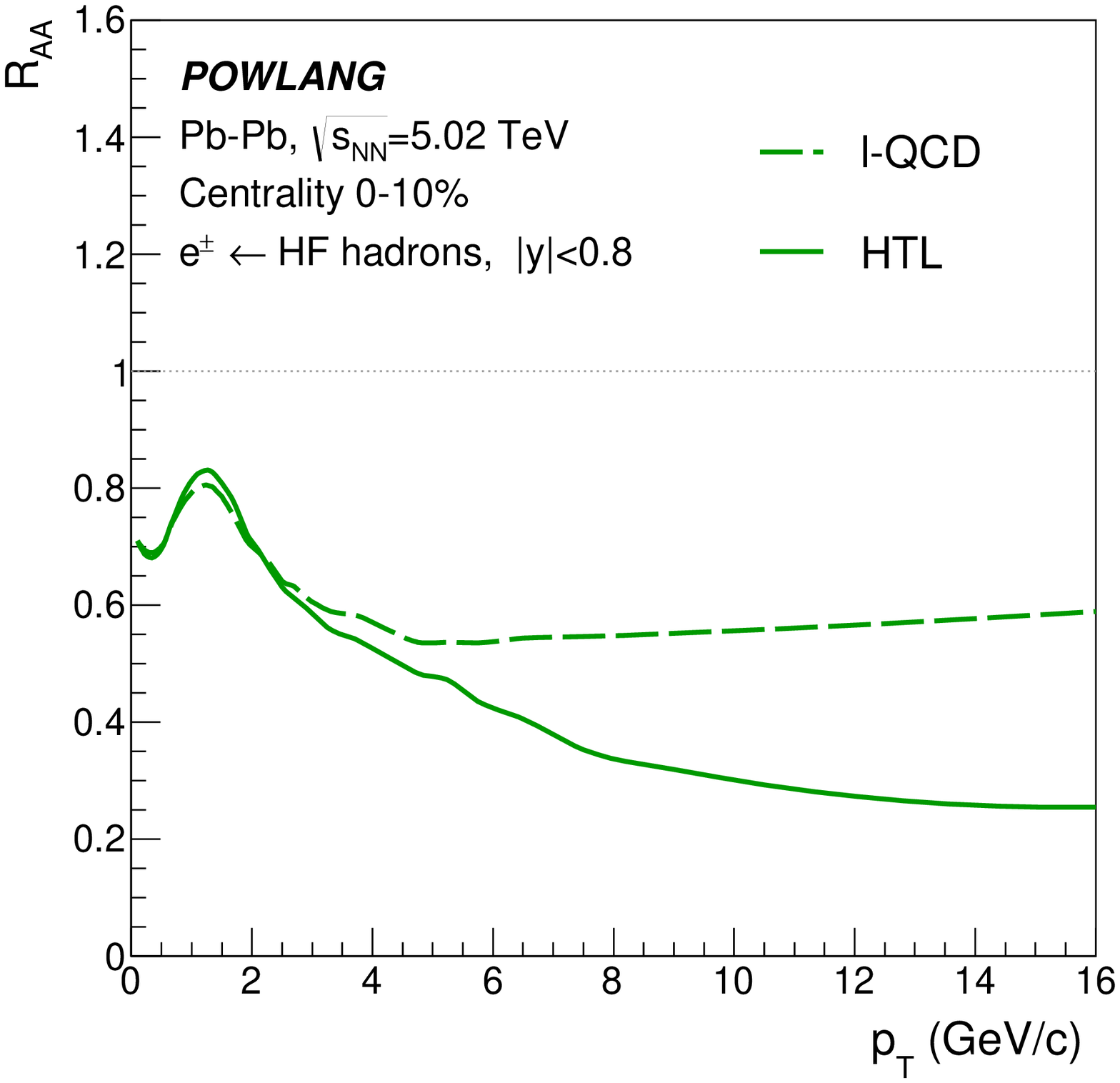}
\includegraphics[clip,height=6cm]{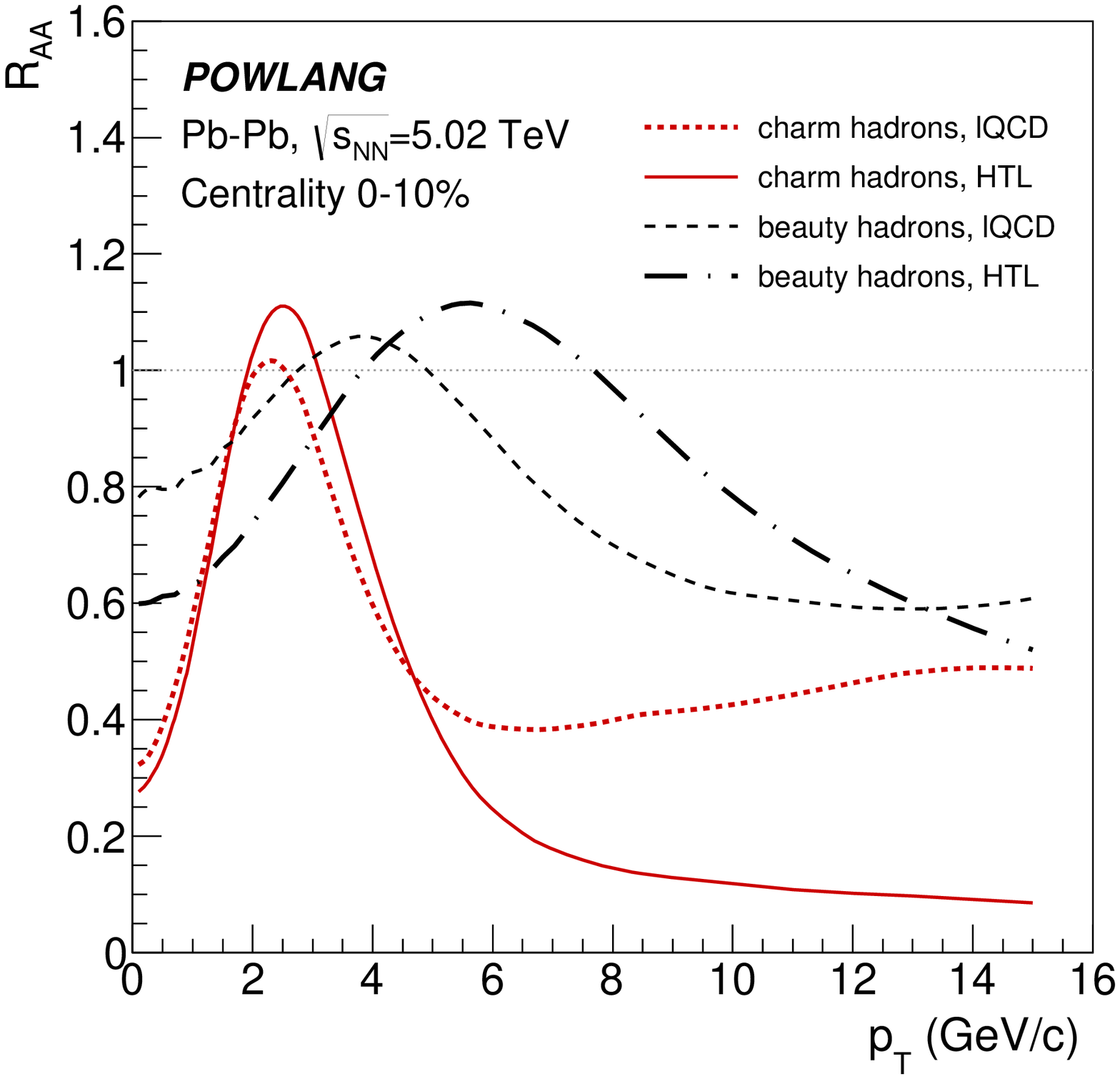}
\caption{The nuclear modification factor of electrons from semi-leptonic decays of heavy-flavour hadrons in central Pb-Pb collisions at $\sqrt{s_{\rm NN}}\!=\!5.02$ TeV. POWLANG results with HTL and l-QCD transport coefficients are compared. In the right panel we display the corresponding results for the $R_{\rm AA}$ of the parent charm and beauty hadrons.}\label{fig:RAA-HFE} 
\end{center}
\end{figure}
Finally, in Fig.~\ref{fig:RAA-HFE} we extend our predictions to the electrons from semi-leptonic decays of charm and beauty hadrons, considering their nuclear modification factor in central Pb-Pb collisions at $\sqrt{s_{\rm NN}}\!=\!5.02$ TeV. We notice that up to $p_T\!\approx\!4$ GeV/c the curves with weak-coupling HTL and lattice-QCD transport coefficients are very similar whereas, as usual, they tend to diverge for higher $p_T$, due to the different momentum dependence of $\kappa_{\perp/\|}$ in the two cases. This behaviour is mainly driven by the one of the parent charm hadrons, displaying very similar momentum distributions at low $p_T$ and a quite different quenching at higher $p_T$, as shown in the right panel of Fig.~\ref{fig:RAA-HFE}. 

\section{Discussion and perspectives}\label{sec:discussion}
In this paper, besides extending to higher center-of-mass energies the predictions of our POWLANG transport model, we tried to study in more detail the development (as a function of the decoupling time from the medium) of the anisotropies in the angular distribution of heavy-flavour particles in relativistic heavy-ion collisions. We considered both the second and third harmonics of the Fourier expansion of the heavy-quark azimuthal distribution at the time of their decoupling, which may occur -- depending on the initial production point -- during the whole lifetime of the fireball arising from the collision of the two nuclei. The second harmonic $v_2$ arises mainly (except in the case of ultra-central events) from the finite impact parameter of the collision, while non-zero values of the third Fourier coefficient $v_3$ are entirely due to event-by-event fluctuations -- in the nucleon positions, but possibly also (not addressed here) in the nucleon structure itself -- giving rise to lumpy initial conditions with a non-vanishing triangular deformation. 
Within the hydrodynamic paradigm, an initial spatial deformation is transferred via the resulting anisotropic pressure gradients to the momentum and angular distribution of the final particles. There is quite a strong consensus in the literature that peculiar features displayed by the soft-hadrons distributions in heavy-ion collisions (flattening of the $p_T$-spectra, baryon-over-meson ratios, non-vanishing value of the various azimuthal harmonics $v_2,v_3,v_4...$) reflect the underlying collective flow of the medium from which they decouple.
Concerning heavy-flavour particles, initially produced off-equilibrium in hard pQCD processes, the measurement of non-vanishing Fourier harmonics of their azimuthal distribution may indicate that the interaction with the crossed medium was sufficiently strong to make them (at least partially) thermalize and take part in the collective flow of the latter. However, before drawing firm conclusions, it is necessary to examine whether other more trivial effects related to the collision geometry may give rise to the same kind of signals. Hence, we performed a systematic analysis of the development of the azimuthal anisotropy of particle distributions through the study of the heavy quarks decoupling from the fireball during the various stages of its expansion; both in the case of $v_2$ and $v_3$ we found a non trivial trend, the final signal arising from the interplay of very different contributions. Heavy quarks were then hadronized, through the fragmentation of color-singlet strings/clusters obtained joining them with thermal partons picked-up from the medium and hence sharing its collective flow. The final results for the elliptic and triangular flow of charmed hadrons in Au-Au and Pb-Pb collisions at RHIC and LHC energies, for various centrality classes, look in quite good agreement with recent experimental data. Actually, as already found in previous studies, the contribution provided by in-medium hadronization turns out to be quite important in moving the results at the partonic level closer to the experimental data. We also checked that our transport calculations, within the same consistent setup, provide reasonable results for the nuclear modification factor of charmed hadrons and electrons from heavy-flavour decays.

Several items would deserve further investigation. Within the same centrality class (in our case identified via the minimum/maximum number of binary nucleon-nucleon collisions) one could examine the effect of eccentricity fluctuations and how they can affect the flow not only of light, but also of heavy-flavour hadrons, by selecting events characterized by a large/small eccentricity (with so-called event-shape engineering techniques). At the same time one could select events characterized by similar eccentricity, but belonging to different centrality classes.
Secondly, we plan to perform transport calculations based on a full (3+1)D modeling of the background medium, dropping the assumption of longitudinal boost-invariance. This, although requiring greater storage and computing resources, will allow us to provide predictions for observables at forward/backward rapidity, so far neglected in our analysis. In particular, this will certainly provide a more realistic description of the background medium in proton-nucleus collisions, in which -- due to the asymmetry of the system -- the assumption of longitudinal boost-invariance is too drastic. The question of the possible hot-medium effects in small-system, also for what concerns heavy flavour, remains in fact open. In a previous publication we showed how our transport setup -- with initial-state nuclear effects, partonic transport in a small QGP droplet and in-medium hadronization -- provides results compatible with the experimental data, within their large systematic error-bars. Nowadays, experimental analysis with larger statistics are in progress and hopefully will provide more differential results for a wider set of observables, allowing one to put tighter constraints on theoretical models and to rule out scenarios not supported by the experimental data. We plan to address the above important items in forthcoming publications. 
\bibliography{paper}

\providecommand{\newblock}{}
\begin{thebibliography}{10}
\expandafter\ifx\csname url\endcsname\relax
  \def\url#1{{\tt #1}}\fi
\expandafter\ifx\csname urlprefix\endcsname\relax\def\urlprefix{URL }\fi
\providecommand{\eprint}[2][]{\url{#2}}

\bibitem{Adare:2006nq}
Adare A {\em et~al.\/} (PHENIX) 2007 {\em Phys. Rev. Lett.\/} {\bf 98} 172301
  (\textit{Preprint} \eprint{nucl-ex/0611018})

\bibitem{ALICE:2012ab}
Abelev B {\em et~al.\/} (ALICE) 2012 {\em JHEP\/} {\bf 09} 112
  (\textit{Preprint} \eprint{1203.2160})

\bibitem{Abelev:2013lca}
Abelev B {\em et~al.\/} (ALICE) 2013 {\em Phys. Rev. Lett.\/} {\bf 111} 102301
  (\textit{Preprint} \eprint{1305.2707})

\bibitem{Chatrchyan:2012np}
Chatrchyan S {\em et~al.\/} (CMS) 2012 {\em JHEP\/} {\bf 05} 063
  (\textit{Preprint} \eprint{1201.5069})

\bibitem{Adamczyk:2014uip}
Adamczyk L {\em et~al.\/} (STAR) 2014 {\em Phys. Rev. Lett.\/} {\bf 113} 142301
  (\textit{Preprint} \eprint{1404.6185})

\bibitem{Adamczyk:2017xur}
Adamczyk L {\em et~al.\/} (STAR) 2017 {\em Phys. Rev. Lett.\/} {\bf 118} 212301
  (\textit{Preprint} \eprint{1701.06060})

\bibitem{Sirunyan:2017plt}
Sirunyan A~M {\em et~al.\/} (CMS) 2017  (\textit{Preprint} \eprint{1708.03497})

\bibitem{Moore:2004tg}
Moore G~D and Teaney D 2005 {\em Phys. Rev.\/} {\bf C71} 064904
  (\textit{Preprint} \eprint{hep-ph/0412346})

\bibitem{Alberico:2011zy}
Alberico W~M, Beraudo A, De~Pace A, Molinari A, Monteno M, Nardi M and Prino F
  2011 {\em Eur. Phys. J.\/} {\bf C71} 1666 (\textit{Preprint}
  \eprint{1101.6008})

\bibitem{Alberico:2013bza}
Alberico W~M, Beraudo A, De~Pace A, Molinari A, Monteno M, Nardi M, Prino F and
  Sitta M 2013 {\em Eur. Phys. J.\/} {\bf C73} 2481 (\textit{Preprint}
  \eprint{1305.7421})

\bibitem{Beraudo:2014boa}
Beraudo A, De~Pace A, Monteno M, Nardi M and Prino F 2015 {\em Eur. Phys. J.\/}
  {\bf C75} 121 (\textit{Preprint} \eprint{1410.6082})

\bibitem{Beraudo:2015wsd}
Beraudo A, De~Pace A, Monteno M, Nardi M and Prino F 2016 {\em JHEP\/} {\bf 03}
  123 (\textit{Preprint} \eprint{1512.05186})

\bibitem{Nahrgang:2014vza}
Nahrgang M, Aichelin J, Bass S, Gossiaux P~B and Werner K 2015 {\em Phys.
  Rev.\/} {\bf C91} 014904 (\textit{Preprint} \eprint{1410.5396})

\bibitem{Prado:2016vgz}
Prado C~A~G, Noronha-Hostler J, Noronha J, Suaide A~A~P, Munhoz M~G and
  Cosentino M~R 2016 {\em {8th International Conference on Hard and
  Electromagnetic Probes of High-energy Nuclear Collisions: Hard Probes 2016
  (HP2016) Wuhan, Hubei, China, September 23-27, 2016}\/} (\textit{Preprint}
  \eprint{1612.05724})
  \urlprefix\url{http://inspirehep.net/record/1504966/files/arXiv:1612.05724.pdf}

\bibitem{Greco:2003vf}
Greco V, Ko C~M and Rapp R 2004 {\em Phys. Lett.\/} {\bf B595} 202--208
  (\textit{Preprint} \eprint{nucl-th/0312100})

\bibitem{Gossiaux:2009mk}
Gossiaux P~B, Bierkandt R and Aichelin J 2009 {\em Phys. Rev.\/} {\bf C79}
  044906 (\textit{Preprint} \eprint{0901.0946})

\bibitem{Nahrgang:2016lst}
Nahrgang M, Aichelin J, Gossiaux P~B and Werner K 2016 {\em Phys. Rev.\/} {\bf
  C93} 044909 (\textit{Preprint} \eprint{1602.03544})

\bibitem{Ravagli:2007xx}
Ravagli L and Rapp R 2007 {\em Phys. Lett.\/} {\bf B655} 126--131
  (\textit{Preprint} \eprint{0705.0021})

\bibitem{He:2011qa}
He M, Fries R~J and Rapp R 2012 {\em Phys. Rev.\/} {\bf C86} 014903
  (\textit{Preprint} \eprint{1106.6006})

\bibitem{He:2014cla}
He M, Fries R~J and Rapp R 2014 {\em Phys. Lett.\/} {\bf B735} 445--450
  (\textit{Preprint} \eprint{1401.3817})

\bibitem{Adam:2015jda}
Adam J {\em et~al.\/} (ALICE) 2016 {\em JHEP\/} {\bf 03} 082 (\textit{Preprint}
  \eprint{1509.07287})

\bibitem{ALICE-PUBLIC-2017-003}
 2017  \urlprefix\url{https://cds.cern.ch/record/2265109}

\bibitem{Zhou:2017ikn}
Zhou L (STAR) 2017  (\textit{Preprint} \eprint{1704.04364})

\bibitem{He:2015hfa}
He L, Edmonds T, Lin Z~W, Liu F, Molnar D and Wang F 2016 {\em Phys. Lett.\/}
  {\bf B753} 506--510 (\textit{Preprint} \eprint{1502.05572})

\bibitem{Xu:2004mz}
Xu Z and Greiner C 2005 {\em Phys. Rev.\/} {\bf C71} 064901 (\textit{Preprint}
  \eprint{hep-ph/0406278})

\bibitem{He:2013zua}
He M, van Hees H, Gossiaux P~B, Fries R~J and Rapp R 2013 {\em Phys. Rev.\/}
  {\bf E88} 032138 (\textit{Preprint} \eprint{1305.1425})

\bibitem{Banerjee:2011ra}
Banerjee D, Datta S, Gavai R and Majumdar P 2012 {\em Phys. Rev.\/} {\bf D85}
  014510 (\textit{Preprint} \eprint{1109.5738})

\bibitem{Francis:2015daa}
Francis A, Kaczmarek O, Laine M, Neuhaus T and Ohno H 2015 {\em Phys. Rev.\/}
  {\bf D92} 116003 (\textit{Preprint} \eprint{1508.04543})

\bibitem{CaronHuot:2008uh}
Caron-Huot S and Moore G~D 2008 {\em JHEP\/} {\bf 02} 081 (\textit{Preprint}
  \eprint{0801.2173})

\bibitem{Xu:2017hgt}
Xu Y, Nahrgang M, Bernhard J~E, Cao S and Bass S~A 2017  (\textit{Preprint}
  \eprint{1704.07800})

\bibitem{DelZanna:2013eua}
Del~Zanna L, Chandra V, Inghirami G, Rolando V, Beraudo A, De~Pace A, Pagliara
  G, Drago A and Becattini F 2013 {\em Eur. Phys. J.\/} {\bf C73} 2524
  (\textit{Preprint} \eprint{1305.7052})

\bibitem{Qiu:2011iv}
Qiu Z and Heinz U~W 2011 {\em Phys. Rev.\/} {\bf C84} 024911 (\textit{Preprint}
  \eprint{1104.0650})

\bibitem{Heinz:2013th}
Heinz U and Snellings R 2013 {\em Ann. Rev. Nucl. Part. Sci.\/} {\bf 63}
  123--151 (\textit{Preprint} \eprint{1301.2826})

\bibitem{ALICE:2011ab}
Aamodt K {\em et~al.\/} (ALICE) 2011 {\em Phys. Rev. Lett.\/} {\bf 107} 032301
  (\textit{Preprint} \eprint{1105.3865})

\bibitem{Aad:2013xma}
Aad G {\em et~al.\/} (ATLAS) 2013 {\em JHEP\/} {\bf 11} 183 (\textit{Preprint}
  \eprint{1305.2942})

\bibitem{Aad:2015lwa}
Aad G {\em et~al.\/} (ATLAS) 2015 {\em Phys. Rev.\/} {\bf C92} 034903
  (\textit{Preprint} \eprint{1504.01289})

\bibitem{ALICE:2016kpq}
Adam J {\em et~al.\/} (ALICE) 2016 {\em Phys. Rev. Lett.\/} {\bf 117} 182301
  (\textit{Preprint} \eprint{1604.07663})

\bibitem{Adare:2013piz}
Adare A {\em et~al.\/} (PHENIX) 2013 {\em Phys. Rev. Lett.\/} {\bf 111} 212301
  (\textit{Preprint} \eprint{1303.1794})

\bibitem{Khachatryan:2015waa}
Khachatryan V {\em et~al.\/} (CMS) 2015 {\em Phys. Rev. Lett.\/} {\bf 115}
  012301 (\textit{Preprint} \eprint{1502.05382})

\bibitem{Aaboud:2017acw}
Aaboud M {\em et~al.\/} (ATLAS) 2017 {\em Eur. Phys. J.\/} {\bf C77} 428
  (\textit{Preprint} \eprint{1705.04176})

\bibitem{Khachatryan:2016txc}
Khachatryan V {\em et~al.\/} (CMS) 2017 {\em Phys. Lett.\/} {\bf B765} 193--220
  (\textit{Preprint} \eprint{1606.06198})

\bibitem{Blok:2017pui}
Blok B, Jäkel C~D, Strikman M and Wiedemann U~A 2017  (\textit{Preprint}
  \eprint{1708.08241})

\bibitem{Dusling:2017aot}
Dusling K, Mace M and Venugopalan R 2017  (\textit{Preprint}
  \eprint{1706.06260})

\bibitem{Dusling:2017dqg}
Dusling K, Mace M and Venugopalan R 2017  (\textit{Preprint}
  \eprint{1705.00745})

\bibitem{Heiselberg:1998es}
Heiselberg H and Levy A~M 1999 {\em Phys. Rev.\/} {\bf C59} 2716--2727
  (\textit{Preprint} \eprint{nucl-th/9812034})

\bibitem{Kolb:2000fha}
Kolb P~F, Huovinen P, Heinz U~W and Heiselberg H 2001 {\em Phys. Lett.\/} {\bf
  B500} 232--240 (\textit{Preprint} \eprint{hep-ph/0012137})

\bibitem{Rapp:2008qc}
Rapp R and van Hees H 2008  (\textit{Preprint} \eprint{0803.0901})

\bibitem{Acharya:2017qps}
Acharya S {\em et~al.\/} (ALICE) 2017  (\textit{Preprint} \eprint{1707.01005})

\bibitem{Sirunyan:2017xss}
Sirunyan A~M {\em et~al.\/} (CMS) 2017  (\textit{Preprint} \eprint{1708.04962})

\end{thebibliography}

\end{document}